\documentclass[a4,12pt]{article}
\usepackage{tikz}
\usepackage[a4paper,left=30mm,right=15mm,top=20mm,bottom=20mm]{geometry}
\usepackage{cite}

\tikzset{block/.style={draw,rounded corners,minimum
width=0.3cm,minimum height=0.5cm}}
\tikzset{fblock/.style={draw,rounded corners,minimum
width=0.3cm,minimum height=0.5cm}}
\tikzset{bblock/.style={draw,minimum width=0.3cm,minimum
height=0.5cm}}
\tikzset{invis/.style={rounded corners,minimum width=0.3cm,minimum
height=0.5cm}}

\def \iz {{{}^1\!\!/\!{}_2}}
\def \tz {{{}^3\!\!/\!{}_2}}
\def \pz {{{}^5\!\!/\!{}_2}}

\title{Massive higher spin supermultiplets unfolded}

\author{M.V. Khabarov${}^{a}$\thanks{maksim.khabarov@ihep.ru},
Yu.M. Zinoviev${}^{ab}$\thanks{Yurii.Zinoviev@ihep.ru}
\\[0.5cm]
\it{\small ${}^a$Institute for High Energy Physics of National
Research Center "Kurchatov Institute"} \\
\it{\small Protvino, Moscow Region, 142281, Russia} \\
\it\small{ ${}^b$Moscow Institute of Physics and Technology (State
University),} \\
\it{\small Dolgoprudny, Moscow Region, 141701, Russia}}

\date{}

\begin{document}

\maketitle

\begin{abstract}
In this paper we construct an unfolded formulation for the massive
higher spin $N=1$ supermultiplets in four dimensional $AdS$ space. We
use the same frame-like gauge invariant multispinor formalism that was
used previously for their Lagrangian formulation. We also consider an
infinite spin limit of such supermultiplets.
\end{abstract}

\thispagestyle{empty}
\newpage
\setcounter{page}{1}
\tableofcontents\pagebreak


\section*{Introduction}

Lagrangian formulation for the massless higher spin supermultiplets
(both on-shell and off-shell, both in flat space and in $AdS$) has
been known for a long time \cite{Cur79,Vas80,KSP93,KS93,KS94}.
However, any attempts to deform massless supermultiplets into the
massive ones lead to the introduction of very complicated higher
derivative corrections to the supertransformations without evident
patterns. Moreover, the higher superspin of the supermultiplets is,
the higher the number of derivatives one has to consider. Even the
usage of the powerful superfield formalism allowed to construct only
a couple of examples with relatively low superspins
\cite{BGPL02,BGLP02}.

For the first time massive arbitrary superspin $N=1$ supermultiplets
in flat four dimensional space were constructed in \cite{Zin07a} using
the gauge invariant formulation for the massive bosonic \cite{Zin01}
and fermionic \cite{Met06} fields. Initial idea was that the massive
supermultiplet can be constructed out of the appropriately chosen set
of the massless ones in the same way as the gauge invariant
description for the massive fields can be constructed out of the
appropriate set of the massless ones. The real picture (in a sense of
the massless limit) appeared to be slightly more complicated, but
anyway the construction was successful.

Later on, the Lagrangian formulation for the higher spin massive
supermultiplets in flat three dimensional space has also been
constructed \cite{BSZ15}, again using the gauge invariant formulation
for massive bosonic and fermionic fields adopted for $d=3$ 
\cite{BSZ12a,BSZ14a}. The correct procedure to deform such
supermultiplets into $AdS_3$ space was not evident form the very
beginning. It so happened that firstly the unfolded formulation has
been
 constructed \cite{BSZ16} based on the results in \cite{Zin15}. After
that the Lagrangian formulation for these supermultiplets in $AdS_3$
has also been completed \cite{BSZ17,BSZ17a}.

Recently, we have managed to construct the Lagrangian formulation for
massive higher spin $N=1$ supermultiplets in $AdS_4$ \cite{BKhSZ19}
using the frame-like gauge invariant formalism \cite{Zin08b} in its
multispinor version adopted for $d=4$. Note that though the
traditional classification of the supermultiplets describes only
massless and massive ones, recently it was shown \cite{GHR18} that in
$AdS_4$ space there exist the non-unitary higher spin supermultiplets
containing partially massless fields. The explicit Lagrangian
formulation for such supermultiplets has been constructed in
\cite{BKhSZ19a}. Note also that the first examples of the infinite
spin supermultiplets in flat space were constructed recently
\cite{Zin17,BKhSZ19b} (see also recent paper \cite{Naj19}). Here again
it was crucial that the gauge invariant formalism used for the
description of massive finite spin fields nicely works for the
infinite spin limit as well
\cite{Met16,Met17,Zin17,KhZ17,Met18,KhZ19}.

The main aim of this paper is to construct unfolded formulation for
the massive higher spin (including infinite spin limit) $N=1$
supermultiplets in $AdS_4$. Recall that the unfolded formulation for
massive higher spin bosons in arbitrary $d \ge 4$ has been constructed
in \cite{PV10}, while such formulation both for bosons as well as
fermions in $AdS_4$ appeared recently in our work \cite{KhZ19}. Note
here that, as far as we know, till now only unfolded formulation for
the scalar supermultiplet was considered \cite{PV10a,MV13}.

The paper is organized as follows. In Section 1 as a simple
illustration of our formalism we provide an unfolded formulation for
the massless $N=1$ supermultiplets. Also, in Section 2 we give a pair
of simple examples for the lower spin massive supermultiplets, namely
the scalar and the vector ones. Section 3 is devoted to the main task
--- construction of the unfolded formulation for the massive arbitrary
superspin $N=1$ supermultiplets. We follow the same strategy as in the
construction of their Lagrangian formulation in \cite{BKhSZ19}.
Namely, first of all we provide the unfolded equations for the massive
bosons and fermions. Then we consider a pair of boson and fermion and
construct supertransformations leaving their unfolded equations
invariant. At last we consider complete supermultiplets containing two
bosons and two fermions and adjust  their parameters so that the
algebra of the supertransformations is closed. Section 4 is devoted to
the infinite spin supermultiplets. 

\section{Massless higher spin supermultiplets}

In this section we provide an unfolded formulation for the massless
higher spin supermultiplets \cite{Cur79,Vas80,KSP93,KS93,KS94} in  the
frame-like multispinor formalism we use later on for the construction
of the massive supermultiplets.

\subsection{Unfolded equations}

Let us briefly recall the unfolded description of massless higher spin
fields (see e.g. \cite{DS14}). To build a system of unfolded equations
for spin-$s$ boson, one needs a set of gauge one-forms
$\Omega^{\alpha(s-1+m)\dot\alpha(s-1-m)}$, $0\le m <s$ and a set of
gauge invariant zero-forms $W^{\alpha(k+s)(k-s)}$, $k\ge s$ with their
conjugates. The field $\Omega^{\alpha(s-1)\dot\alpha(s-1)}$ is the
physical one. The gauge transformations for the one-forms are:
\begin{eqnarray}
\delta\Omega^{\alpha(s-1)\dot\alpha(s-1)} &=& 
D\eta^{\alpha(s-1)\dot\alpha(s-1)}
+ e^{\alpha}{}_{\dot\beta} \eta^{\alpha(s-2)\dot\alpha(s-1)\dot\beta}
+ e_{\beta}{}^{\dot\alpha} \eta^{\alpha(s-1)\beta\dot\alpha(s-2)}
\nonumber \\
\delta\Omega^{\alpha(s-1+m)\dot\alpha(s-1-m)} &=& 
D\eta^{\alpha(s-1+m)\dot\alpha(s-1-m)}
+ \lambda^2e^{\alpha}{}_{\dot\beta}
\eta^{\alpha(s-2+m)\dot\alpha(s-m-1)\dot\beta}
\nonumber \\
&& + e_{\beta}{}^{\dot\alpha} 
\eta^{\alpha(s+m-1)\beta\dot\alpha(s-2-m)}, \qquad 0<m<s-1
 \\
\delta\Omega^{\alpha(2s-2)} &=& D\eta^{\alpha(2s-2)}
+ \lambda^2e^{\alpha}{}_{\dot\alpha} 
\eta^{\alpha(2s-3)\dot\alpha}
\nonumber
\end{eqnarray}
A set of gauge invariant two-forms - "curvatures" - can be build from
these one-forms:
\begin{eqnarray}
\mathcal{R}^{\alpha(s-1)\dot\alpha(s-1)} &=&
D\Omega^{\alpha(s-1)\dot\alpha(s-1)}
+ e^{\alpha}{}_{\dot\beta} 
\Omega^{\alpha(s-2)\dot\alpha(s-1)\dot\beta}
+ e_{\beta}{}^{\dot\alpha} \Omega^{\alpha(s-1)\beta\dot\alpha(s-2)}
\nonumber \\
\mathcal{R}^{\alpha(s-1+m)\dot\alpha(s-1-m)} &=&
D\Omega^{\alpha(s-1+m)\dot\alpha(s-1-m)}
+ \lambda^2e^{\alpha}{}_{\dot\beta}
\Omega^{\alpha(s-2+m)\dot\alpha(s-m-1)\dot\beta}
\nonumber \\
&& + e_{\beta}{}^{\dot\alpha}
\Omega^{\alpha(s+m-1)\beta\dot\alpha(s-2-m)}, \qquad 0<m<s-1
 \\
\mathcal{R}^{\alpha(2s-2)} &=& D\Omega^{\alpha(2s-2)}
+ \lambda^2e^{\alpha}{}_{\dot\alpha} \Omega^{\alpha(2s-3)\dot\alpha}
\nonumber
\end{eqnarray}
The system of unfolded equations then can be split into the three
parts. The first one is the zero-curvature conditions (analogue of the
zero torsion condition in gravity):
\begin{equation}
\mathcal{R}^{\alpha(s-1+m)\dot\alpha(s-1-m)} = 0,  \qquad m<s-1
\end{equation}
while the second one connects the one-forms and zero-forms sectors:
\begin{equation}
\mathcal{R}^{\alpha(2s-2)} = -2E_{\beta(2)}W^{\alpha(2s-2)\beta(2)},
\end{equation}
and the third one contains gauge invariant zero-forms only:
\begin{eqnarray}
0 &=& DW^{\alpha(2s)} + e_{\beta\dot\alpha}
W^{\alpha(2s)\beta\dot\alpha},
\nonumber \\
0 &=& DW^{\alpha(2s+m)\dot\alpha(m)} + e_{\beta\dot\beta}
W^{\alpha(2s+m)\beta\dot\alpha(m)\dot\beta} + 
\lambda^2e^{\alpha\dot\alpha}
W^{\alpha(2s+m-1)\dot\alpha(m-1)},\quad m>0
\end{eqnarray}
The unfolded equations can be regarded as a chain of equations of the
form $DA_i = eA_{i+1}+O(\lambda^2)$. This means that the field
$A_{i+1}$ is a parametrization of the derivatives of $A_i$, which do
not vanish on-shell, up to the gauge transformations.

In a similar fashion, the description of the massless fermion with
spin ${\tilde{s}}=s+\iz$ is built. One needs a set of gauge one-forms
$\Psi^{\alpha({\tilde{s}}-1+m)\dot\alpha({\tilde{s}}-1-m)}$, 
$\iz\le m < {\tilde{s}}$ and a set of gauge invariant zero-forms
$Y^{\alpha(k+{\tilde{s}})(k-{\tilde{s}})}$, $k\ge {\tilde{s}}$ with
their conjugates (where indices $k,m$ are half-integer). Here, the
pair of fields 
$\Psi^{\alpha({\tilde{s}}-1\pm\iz)\dot\alpha({\tilde{s}}-1\mp\iz)}$
play the role of the physical ones. Similarly, a set of curvatures
$\mathcal{F}^{\alpha({\tilde{s}}-1+m)\dot\alpha({\tilde{s}}-1-m)}$ can
be constructed. The expressions for the gauge transformations and
curvatures are similar to the bosonic case (up to the change
$s\to\tilde{s}$, $\Omega\to\Psi$ and half-integer $m$), with the only
exception being the case $m=\iz$:
\begin{eqnarray}
\delta\Psi^{\alpha({\tilde{s}}-\iz)\dot\alpha(s-\tz)}
&=& D\eta^{\alpha({\tilde{s}}-\iz)\dot\alpha({\tilde{s}}-\tz)}
+ e^{\alpha}{}_{\dot\beta}
\eta^{\alpha({\tilde{s}}-\iz)\dot\alpha({\tilde{s}}-\pz)\dot\beta}
\nonumber \\
&& + \epsilon\lambda e_{\beta}{}^{\dot\alpha}
\eta^{\alpha({\tilde{s}}-\pz)\beta\dot\alpha({\tilde{s}}-\iz)}
 \\
\mathcal{F}^{\alpha({\tilde{s}}-\iz)\dot\alpha(s-\tz)}
&=& D\Psi^{\alpha({\tilde{s}}-\iz)\dot\alpha({\tilde{s}}-\tz)}
+ e^{\alpha}{}_{\dot\beta}
\Psi^{\alpha({\tilde{s}}-\iz)\dot\alpha({\tilde{s}}-\pz)\dot\beta}
\nonumber \\
&& + \epsilon\lambda e_{\beta}{}^{\dot\alpha}
\Psi^{\alpha({\tilde{s}}-\pz)\beta\dot\alpha({\tilde{s}}-\iz)} 
\end{eqnarray}
In case of AdS space, the parameter $\epsilon=\pm 1$ here corresponds
to the choice of the sign of mass-like terms. In the flat space,
however, this parameter is arbitrary for the massless particle. With
the gauge forms encapsulated in curvatures, the unfolded equations
reproduce the exact form of the bosonic ones:
\begin{eqnarray}
0 &=& 
\mathcal{F}^{\alpha({\tilde{s}}-1+m)\dot\alpha({\tilde{s}}-1-m)}, 
\qquad m<{\tilde{s}}-1
\nonumber \\
0 &=& \mathcal{F}^{\alpha(2{\tilde{s}}-2)} -
2E_{\beta(2)}Y^{\alpha(2{\tilde{s}}-2)\beta(2)},
\nonumber \\
0 &=& DY^{\alpha(2{\tilde{s}})} +
e_{\beta\dot\alpha}Y^{\alpha(2{\tilde{s}})\beta\dot\alpha},
\\
0 &=& DY^{\alpha(2{\tilde{s}}+m)\dot\alpha(m)} +
e_{\beta\dot\beta}Y^{\alpha(2{\tilde{s}}+m)\beta\dot\alpha(m)\dot\beta}
\nonumber \\
&& + \lambda^2e^{\alpha\dot\alpha}
Y^{\alpha(2{\tilde{s}}+m-1)\dot\alpha(m-1)}, \quad m \ge\iz \nonumber
\end{eqnarray}
Note once again that numbers $k,m$ are half-integers here.

Now we construct the massless supermultiplets. First, we introduce a
supertransformation parameter $\zeta^{\alpha}$ with its hermitian
conjugate $\zeta^{\dot\alpha}$ which obeys $D\zeta^{\alpha}=-\lambda
e^{\alpha}{}_{\dot\alpha}\zeta^{\dot\alpha}$ (similarly for the
hermitian conjugate).  In the supermultiplet, the spins of boson and
fermion are connected by the relation $\tilde{s}-s=\pm \iz$, so there
are two possibilities.

\subsection{Half-integer superspin}

Our task here to construct supertransformations transforming bosonic
equations into the fermionic ones and vice versa. It is natural to
begin with the gauge invariant zero-forms because they form a closed
sector. The most general ansatz for their supertransformations is
rather simple:
\begin{eqnarray}
\delta W^{\alpha(k+s)\dot\alpha(k-s)} &=&
\delta^{-0}_k Y^{\alpha(k+s-1)\dot\alpha(k-s)} \zeta^{\alpha}
+ \delta^{0+}_k Y^{\alpha(k+s)\dot\alpha(k-s)\dot\beta}
\zeta_{\dot\beta},
\nonumber \\
\delta Y^{\alpha(k+s-1)\dot\alpha(k-s)} &=&
\tilde{\delta}^{+0}_{k-\iz}
W^{\alpha(k+s-1)\beta\dot\alpha(k-s)}\zeta_{\beta}
+ \tilde{\delta}^{0-}_{k-\iz}
W^{\alpha(k+s-1)\dot\alpha(k-s-1)}\zeta^{\dot\alpha}
\end{eqnarray}
where all the coefficients are in general complex. The solution  for
these coefficients turns out to be also simple:
\begin{equation}
\delta^{0+}_k = C_b, \qquad 
\delta^{-0}_k = \lambda C_b, \qquad
\tilde{\delta}^{+0}_{k-\iz} = C_f, \qquad 
\tilde{\delta}^{0-}_{k-\iz} = \lambda C_f
\end{equation}
where $C_b$ and $C_f$ are two independent parameters (see below).

Similarly, the supertransformations for the gauge one-forms (except a
pair of the highest ones) look like:
\begin{eqnarray}
\delta\Omega^{\alpha(s-1+m)\dot\alpha(s-1-m)} &=&
\gamma^{-0}_m \Psi^{\alpha(s-2+m)\dot\alpha(s-1-m)} \zeta^{\alpha}
 + \gamma^{0-}_m \Psi^{\alpha(s-1+m)\dot\alpha(s-2-m)}
\zeta^{\dot\alpha},
\nonumber \\
\delta\Psi^{\alpha(s-2+m)\dot\alpha(s-1-m)} &=&
\tilde{\gamma}^{+0}_{m-\iz}
\Omega^{\alpha(s-2+m)\beta\dot\alpha(s-1-m)} \zeta_{\beta}
+ \tilde{\gamma}^{0+}_{m-\iz}
\Omega^{\alpha(s-2+m)\dot\alpha(s-1-m)\dot\beta} \zeta_{\dot\beta}
\end{eqnarray}
This gives the following solution for the coefficients with $m > 0$:
\begin{equation}
\label{mlhssm_coeff1}
\gamma_m^{0-} = C, \qquad
\gamma_m^{-0} = \lambda C, \qquad
\tilde{\gamma}^{+0}_{m+\iz} = \tilde{C}, \qquad
\tilde{\gamma}^{0+}_{m+\iz} = \lambda \tilde{C}
\end{equation}
where $C$ and $\tilde{C}$ are also independent. For $m=0$, we obtain 
$\gamma_{0}^{0-}=-C$, while for $m<0$, we have:
\begin{eqnarray}
\label{mlhssm_conj1}
\gamma^{-0}_{m}=\epsilon\gamma^{0-}_{-m},\qquad
\gamma^{0-}_{m}=\epsilon\gamma^{-0}_{-m},\qquad
\tilde{\gamma}^{+0}_{m}=\epsilon\tilde{\gamma}^{0+}_{-m},\qquad
\tilde{\gamma}^{0+}_{m}=\epsilon\tilde{\gamma}^{+0}_{-m}.
\end{eqnarray}
At last, we have to consider two highest one-forms 
$\Omega^{\alpha(2s-2)}$ and $\Psi^{\alpha(2s-3)}$ (with their
conjugates) because their equations connect the two sectors. The
ansatz for the supertransformations is now:
\begin{eqnarray}
\delta\Omega^{\alpha(2s-2)} &=&
\nu e_{\alpha\dot\alpha} Y^{\alpha(2s-1)} \zeta^{\dot\alpha}
+ \gamma^{0-}_{s-1} \Psi^{\alpha(2s-3)} \zeta^{\alpha},
\nonumber \\
\delta \Psi^{\alpha(2s-3)} &=& \tilde{\gamma}^{+0}_{s-\tz}
\Omega^{\alpha(2s-3)\beta} \zeta_\beta + \tilde{\gamma}^{0+}_{s-\tz}
\Omega^{\alpha(2s-3)\dot\alpha} \zeta_{\dot\alpha}
\end{eqnarray}
and this provides  the relations on the parameters from the two
sections and fixes the only remaining coefficient:
\begin{equation}
C_b = C, \qquad C_f = \tilde{C}, \qquad \nu = \frac{C}{2}
\end{equation}

The hermiticity requires that $C=-\epsilon C^*$,
$\tilde{C}=\epsilon\tilde{C}^*$. Then, either $C$ is imaginary and
$\epsilon=1$ or $C$ is real and $\epsilon=-1$. The sign of $C^2$
determines the parity of the boson: it is even if $C^2>0$ and odd if
$C^2<0$. Thus, bosonic parity and fermionic mass terms sign are
connected. It is impossible to link $C$ and $\tilde{C}$ by considering
unfolded equations only. However, these constants can be connected if
one requires that the sum of their Lagrangians is invariant under the
supertransformations. If one chooses the normalization of the
Lagrangians as in \cite{KhZ19}, it turns out that:
\begin{eqnarray}
C = 4i\epsilon(s-1) \tilde{C}
\end{eqnarray}
Finally, we evaluate a commutator of two supertransformations to show
that the superalgebra is indeed closed. Consider, for instance, the
field $\Omega^{\alpha(s-1+m)\dot\alpha(s-1-m)}$ for $m>0$. We obtain:
\begin{eqnarray}
[\delta_1,\delta_2] \Omega^{\alpha(s-1+m)\dot\alpha(s-1-m)}
&=& C\tilde{C} \big[\lambda
\Omega^{\alpha(s-2+m)\beta\dot\alpha(s-1-m)} \eta_{\beta}{}^{\alpha} +
\lambda \Omega^{\alpha(s-1+m)\dot\beta\dot\alpha(s-2-m)}
\eta_{\dot\beta}{}^{\dot\alpha}
\nonumber \\
&& \quad + \lambda^2\Omega^{\alpha(s-1+m)\beta\dot\alpha(s-2-m)}
\xi_{\beta}{}^{\dot\alpha} +
\Omega^{\alpha(s-2+m)\dot\beta\dot\alpha(s-1-m)}
\xi^{\alpha}{}_{\dot\beta} \big]
\end{eqnarray}
where
\begin{equation}
\xi^{\alpha\dot\alpha} =
{\zeta_2}^{\alpha}{\zeta_1}^{\dot\alpha}-{\zeta_1}^{\alpha}{\zeta_2}^{\dot\alpha}, \qquad
\eta^{\alpha(2)} = 2{\zeta_2}^{\alpha}{\zeta_1}^{\alpha}, \qquad
\eta^{\dot\alpha(2)} = 2{\zeta_2}^{\dot\alpha}{\zeta_1}^{\dot\alpha}.
\end{equation}
and it is indeed a combination of pseudotranslations and Lorentz
transformations. The expressions for other fields are similar.

Now let us consider the flat space case. Contrary to the $AdS$ case,
the equations for the coefficients with positive and negative $m$ 
fall into two independent subsystems so that we loose the hermiticity
conditions on the parameters $C$ and $\tilde{C}$. The non-zero
coefficients now are:
\begin{eqnarray}
\delta_{k}^{0+}&=& C, \qquad
\tilde{\delta}_{k-\iz}^{+0}=\tilde{C}, \qquad
\nu=\frac{C}{2},
\nonumber \\
\gamma_{m}^{0-}&=&C, \qquad
\tilde{\gamma}^{+0}_{m+\iz}=\tilde{C}, \qquad m\ge 0,
\\
\gamma_{m}^{-0}&=&C^*, \qquad
\tilde{\gamma}^{0+}_{m-\iz}=\tilde{C}^*, \quad m\le 0. 
\nonumber
\end{eqnarray}
To fix the phases of the coefficients $C$ and $\tilde{C}$, one has to
consider a commutator of two supertransformations.  Consider, for
instance, field $\Omega^{\alpha(s-1+m)\dot\alpha(s-1-m)}$, $m>0$. The
commutator of the supertransformations parametrized by
${\zeta_1}^\alpha$, ${\zeta_2}^\alpha$ is:
\begin{equation}
[\delta_1,\delta_2]\Omega^{\alpha(s-1+m)\dot\alpha(s-1-m)} = 
C \tilde{C} \Omega^{\alpha(s+m)\dot\alpha(s-m-2)}
\xi_{\alpha}{}^{\dot\alpha}.
\end{equation}
The hermiticity requires that $C \tilde{C}$ is imaginary. With the
requirement that the sum of the Lagrangians is invariant, a stronger
condition can be obtained:
\begin{eqnarray}
C=4i(s-1) \tilde{C}^*
\end{eqnarray}

\subsection{Integer superspin}

Again, we consider the $AdS$ case first. As in the previous case we
begin with the sector of the gauge invariant zero-forms. In this case
the most general ansatz for the supertransformations is:
\begin{eqnarray}
\delta W^{\alpha(k+s)\dot\alpha(k-s)} &=&
\delta^{+0}_{k}Y^{\alpha(k+s)\beta\dot\alpha(k-s)}\zeta_{\beta}
+ \delta^{0-}_{k}Y^{\alpha(k+s)\dot\alpha(k-s-1)}\zeta^{\dot\alpha},
\nonumber \\
\delta Y^{\alpha(k+s+1)\dot\alpha(k-s)} &=&
\tilde{\delta}^{-0}_{k+\iz}W^{\alpha(k+s)\dot\alpha(k-s)}\zeta^{\alpha}
+\tilde{\delta}^{0+}_{k+\iz}W^{\alpha(k+s+1)\dot\alpha(k-s)\dot\beta}\zeta_{\dot\beta},
\end{eqnarray}
where all coefficients are in general complex. The invariance of the
unfolded equations under these supertransformations leads to:
\begin{equation}
\delta_{k}^{+0} = C_b, \qquad 
\delta_{k}^{0-} = \lambda C_b, \qquad
\tilde{\delta}_{k+\iz}^{0+}= C_f, \qquad
\tilde{\delta}_{k+\iz}^{-0}=\lambda C_f.
\end{equation}
where $C_b$ and $C_f$ are two independent parameters.

Now let us consider a sector of gauge one-forms (except two highest
ones $\Omega^{\alpha(2s-2)}$ and $\Psi^{\alpha(2s-1)}$ with their
conjugates). Here the ansatz for the supertransformations looks like:
\begin{eqnarray}
\delta\Omega^{\alpha(s-1+m)\dot\alpha(s-1-m)}&=&
\gamma^{+0}_{m}
\Psi^{\alpha(s-1+m)\beta\dot\alpha(s-1-m)}\zeta_{\beta} +
\gamma^{0+}_{m}
\Psi^{\alpha(s-1+m)\dot\alpha(s-1-m)\dot\beta} \zeta_{\dot\beta},
\nonumber \\
\delta\Psi^{\alpha(s+m)\dot\alpha(s-1-m)} &=&
\tilde{\gamma}^{-0}_{m+\iz}
\Omega^{\alpha(s-1+m)\dot\alpha(s-1-m)}\zeta^{\alpha}
 + \tilde{\gamma}^{0-}_{m+\iz}
\Omega^{\alpha(s+m)\dot\alpha(s-2-m)}\zeta^{\dot\alpha},
\end{eqnarray}
and the solution gives us two additional independent parameters:
\begin{equation}
\gamma^{+0}_{m} = C, \qquad
\gamma^{0+}_{m} = \lambda C, \qquad
\tilde{\gamma}_{m+\iz}^{0-}=\tilde{C}, \qquad
\tilde{\gamma}_{m+\iz}^{-0} = \lambda \tilde{C} \qquad m > 0.
\end{equation}
For $m=0$, we obtain $\gamma_{0}^{0+}=-C$, while for $m<0$, we have:
\begin{eqnarray}
\gamma^{-0}_{m}=\epsilon\gamma^{0-}_{-m},\qquad
\gamma^{0-}_{m}=\epsilon\gamma^{-0}_{-m},\qquad
\tilde{\gamma}^{+0}_{m}=\epsilon\tilde{\gamma}^{0+}_{-m}, \qquad
\tilde{\gamma}^{0+}_{m}=\epsilon\tilde{\gamma}^{+0}_{-m}
\end{eqnarray}

At last, we consider supertransformations for the remaining one-forms:
\begin{eqnarray}
\delta \Omega^{\alpha(2s-2)} &=& \gamma^{+0}_{s-1}
\Psi^{\alpha(2s-2)\beta} \zeta_\beta + \gamma^{0+}_{s-1}
\Psi^{\alpha(2s-2)\dot\beta} \zeta_{\dot\beta}, 
\nonumber \\
\delta\Psi^{\alpha(2s-1)} &=&
\tilde{\nu} e_{\beta\dot\alpha} W^{\alpha(2s-1)\beta} 
\zeta^{\dot\alpha} + \tilde{\gamma}^{0-}_{s-\iz}
\Omega^{\alpha(2s-2)}\zeta^{\alpha}
\end{eqnarray}
which gives us the relations between the parameters of the two sectors
and determines the only remaining one:
\begin{equation}
C_b = C, \qquad C_f = \tilde{C}, \qquad
\tilde{\nu} = \frac{\tilde{C}}{2}.
\end{equation}

Again, this gives $C=-\epsilon C^*$, $\tilde{C}=\epsilon\tilde{C}^*$
together with the hermiticity requirement. Hence, the boson has the
parity opposite to $\epsilon$, similarly to the half-integer superspin
case. By considering the unfolded equations only, the only thing one
can establish is that the product of the parameters $C$ and 
$\tilde{C}$ must be imaginary. The constants $C$ and $\tilde{C}$ can
be linked by requirement that the sum of bosonic and fermionic
Lagrangians is invariant under the supertransformations:
\begin{eqnarray}
(s-1)C = 4i\epsilon \tilde{C}
\end{eqnarray}

The expression for the commutator of two supertransformations
parametrized by ${\zeta_1}^\alpha$ and ${\zeta_2}^\alpha$ is the same
as in the previous case:
\begin{eqnarray}
[\delta_1,\delta_2] \Omega^{\alpha(s-1+m)\dot\alpha(s-1-m)}
&=& C\tilde{C}
\big[\lambda\Omega^{\alpha(s-2+m)\beta\dot\alpha(s-1-m)}
\eta_{\beta}{}^{\alpha} + \lambda
\Omega^{\alpha(s-1+m)\dot\beta\dot\alpha(s-2-m)}
\eta_{\dot\beta}{}^{\dot\alpha}
\nonumber \\
&& + \lambda^2\Omega^{\alpha(s-1+m)\beta\dot\alpha(s-2-m)}
\xi_{\beta}{}^{\dot\alpha} + 
\Omega^{\alpha(s-2+m)\dot\beta\dot\alpha(s-1-m)}
\xi^{\alpha}{}_{\dot\beta} \big],
\end{eqnarray}

In the flat space, the invariance of the unfolded equations does not
fix the phases of $C$ and $\tilde{C}$, so that the solution for the
coefficients is:
\begin{eqnarray}
\delta_{k}^{+0} &=& C, \qquad
\tilde{\delta}_{k+\iz}^{0+} = \tilde{C}, \qquad
\tilde{\nu}=\frac{\tilde{C}}{2},
\nonumber \\
\gamma^{+0}_{m} &=& C, \qquad
\tilde{\gamma}_{m+\iz}^{0-} = \tilde{C}, \qquad m\ge 0,
\\
\gamma^{+0}_{m}&=&C^*, \qquad
\tilde{\gamma}_{m+\iz}^{0-}=\tilde{C}^*, \qquad m < 0. 
\nonumber
\end{eqnarray}
In this case the requirement that the $C\tilde{C}$ is imaginary
follows only from the commutator of two supertransformations. A
stronger relation
\begin{eqnarray}
(s-1)C = 4i \tilde{C}^*
\end{eqnarray}
can still be obtained from the invariance of the sum of the two
Lagrangians.

\section{Low spins examples}

In this section we present two simplest examples of the massive $N=1$
supermultiplets: a scalar and a vector ones.

\subsection{Unfolded equations}

First of all we need the unfolded equations for massive spin 1,
spin $\iz$ and spin 0 fields.
{\bf Massive vector} In this case the unfolded formulations requires
three infinite chains of the zero-forms: 
$W^{\alpha(k+m)\dot\alpha(k-m)}$, $k \ge 1$, $m = \pm 1,0$
corresponding to the three physical helicities $\pm 1,0$. The most
general (up to the normalization) ansatz has the form: 
\begin{eqnarray}
0 &=& D W^{\alpha(k+1)\dot\alpha(k-1)} + e_{\beta\dot\beta}
W^{\alpha(k+1)\beta\dot\alpha(k-1)\dot\beta} + \beta^{-+}_{k,1}
e^\alpha{}_{\dot\beta} W^{\alpha(k)\dot\alpha(k-1)\dot\beta} +
\beta^{--}_{k,k} e^{\alpha\dot\alpha} W^{\alpha(k)\dot\alpha(k-2)}
\nonumber 
\\
0 &=& D W^{\alpha(k)\dot\alpha(k)} + e_{\beta\dot\beta}
W^{\alpha(k)\beta\dot\alpha(k)\dot\beta} + \beta^{-+}_{k,0}
e^\alpha{}_{\dot\beta} W^{\alpha(k-1)\dot\alpha(k)\dot\beta} \nonumber
\\
 && + \beta^{+-}_{k,0} e_\beta{}^{\dot\alpha} 
W^{\alpha(k)\beta\dot\alpha(k-1)} + \beta^{--}_{k,0}
e^{\alpha\dot\alpha} W^{\alpha(k-1)\dot\alpha(k-1)} 
\\
0 &=& D W^{\alpha(k-1)\dot\alpha(k+1)} + e_{\beta\dot\beta}
W^{\alpha(k-1)\beta\dot\alpha(k+1)\dot\beta} + \beta^{+-}_{k,k}
e_\beta{}^{\dot\alpha} W^{\alpha(k-1)\beta\dot\alpha(k)} +
\beta^{--}_{k,1} e^{\alpha\dot\alpha} W^{\alpha(k-2)\dot\alpha(k)}
\nonumber
\end{eqnarray}
The self-consistency of these equations leads to the following
solutions for the coefficients:
\begin{eqnarray}
\beta^{+-}_{k,0} &=& \beta^{-+}_{k,0} = \frac{1}{k(k+1)} \nonumber
\\
\beta^{+-}_{k,1} &=& \beta^{-+}_{k,1} = \frac{2m^2}{(k+1)(k+2)}
\nonumber
\\
\beta^{--}_{k,1} &=& - \frac{1}{k(k+1)} [m^2 - k(k+1)\lambda^2] 
\\
\beta^{--}_{k,0} &=& - \frac{(k-1)(k+2)}{k^2(k+1)^2}
[m^2 - k(k+1)\lambda^2] \nonumber
\end{eqnarray}

As is well known, in the flat Minkowski space all the members of the
supermultiplet must have equal masses. But in $AdS$ space, as it has
been shown in \cite{BKhSZ19}, there must be a small splitting between
the bosonic and fermionic masses of the order of the cosmological
constant. For the lower spins we consider in this section, the bosonic
mass $m$ and the fermionic one $\tilde{m}$ must satisfy:
\begin{equation}
m^2 = \tilde{m}(\tilde{m} \pm \lambda)
\end{equation}
In this case the $\beta$-functions take the form:
\begin{eqnarray}
\beta^{+-}_{k,0} &=& \beta^{-+}_{k,0} = \frac{1}{k(k+1)} \nonumber
\\
\beta^{+-}_{k,1} &=& \beta^{-+}_{k,1} = 
\frac{2\tilde{m}(\tilde{m}\pm\lambda)}{(k+1)(k+2)} 
\\
\beta^{--}_{k,1} &=& - \frac{1}{k(k+1)} 
[\tilde{m} \pm (k+1)\lambda][\tilde{m} \mp k\lambda] \nonumber 
\\
\beta^{--}_{k,0} &=& - \frac{(k-1)(k+2)}{k^2(k+1)^2}
[\tilde{m} \pm (k+1)\lambda][\tilde{m} \mp k\lambda] \nonumber
\end{eqnarray}
It is this factorization of the $\beta^{--}$-functions that appears to
be crucial  for the construction of the supermultiplets in what
follows. \\
{\bf Massive spinor} In this case there are two physical helicities
$\pm 1/2$ and we need a pair of (conjugated) chains of the zero-forms
$Y^{\alpha(k+1)\dot\alpha(k)}$, $Y^{\alpha(k)\dot\alpha(k+1)}$,
$k \ge 0$. We choose the following ansatz for the unfolded equations:
\begin{eqnarray}
0 &=& D Y^{\alpha(k+1)\dot\alpha(k)} + e_{\beta\dot\beta}
Y^{\alpha(k+1)\beta\dot\alpha\dot\beta} + \tilde{\beta}^{-+}_k
e^\alpha{}_{\dot\beta} Y^{\alpha(k)\dot\alpha(k)\dot\beta} +
\tilde{\beta}^{--}_k e^{\alpha\dot\alpha} Y^{\alpha(k)\dot\alpha(k-1)}
\nonumber 
\\
0 &=& D Y^{\alpha(k)\dot\alpha(k+1)} + e_{\beta\dot\beta}
Y^{\alpha(k)\beta\dot\alpha(k+1)\dot\beta} + \tilde{\beta}^{+-}_k
e_\beta{}^{\dot\alpha} Y^{\alpha(k)\beta\dot\alpha(k)} +
\tilde{\beta}^{--}_k e^{\alpha\dot\alpha} Y^{\alpha(k-1)\dot\alpha(k)}
\end{eqnarray}
The self-consistency of these equations requires:
\begin{eqnarray}
\tilde{\beta}^{+-}_k &=& \tilde{\beta}^{-+}_k = 
\frac{\epsilon \tilde{m}}{(k+1)(k+2)}, \qquad \epsilon = \pm 1
\nonumber
\\
\tilde{\beta}^{--}_k &=& - \frac{1}{(k+1)^2} 
[\tilde{m}{}^2 - (k+1)^2\lambda^2]
\end{eqnarray}
Note that in what follows we always assume that the fermionic masses
are positive and take into account the two possible signs of the
$\tilde{\beta}^{+-}$ (which also play an important role in our
construction) using the parameter $\epsilon = \pm 1$. \\
{\bf Massive scalar} In this case we have one chain of the zero-forms
only with the unfolded equations:
\begin{equation}
0 = D W^{\alpha(k)\dot\alpha(k)} + e_{\beta\dot\beta}
W^{\alpha(k)\beta\dot\alpha(k)\dot\beta} + \beta^{--}_k
e^{\alpha\dot\alpha} W^{\alpha(k-1)\dot\alpha(k-1)}
\end{equation}
where
$$
\beta^{--}_k = - \frac{1}{k(k+1)} [m_0{}^2 - k(k+1)\lambda^2]
$$
As in the spin 1 case, the factorization of the $\beta^{--}$ function
is achieved at $m_0{}^2 = \tilde{m}(\tilde{m} \pm \lambda)$:
\begin{equation}
\beta^{--}_k = - \frac{1}{k(k+1)} [\tilde{m} \pm (k+1)\lambda]
[\tilde{m} \mp k\lambda]
\end{equation}

\subsection{Scalar supermultiplet}

In the flat case such supermultiplet was considered in
\cite{PV10a,MV13}. We begin with a one pair of spinor and scalar
fields. Our first task is to find supertransformations such that the
variations of the fermionic equations be proportional to the bosonic
ones and vice versa. \\
{\bf Supertransformations for spinor} We choose the following ansatz
for the supertransformations where the coefficients are in general
complex:
\begin{eqnarray}
\delta Y^{\alpha(k+1)\dot\alpha(k)} &=& 
\delta^{-0}_k W^{\alpha(k)\dot\alpha(k)} \zeta^\alpha
+ \delta^{0+}_k W^{\alpha(k+1)\dot\alpha(k)\dot\beta} 
\zeta_{\dot\beta} \nonumber 
\\
\delta Y^{\alpha(k)\dot\alpha(k+1)} &=& 
\delta^{+0}_k W^{\alpha(k)\beta\dot\alpha(k+1)} \zeta_\beta 
+ \delta^{0-}_k W^{\alpha(k)\dot\alpha(k)} \zeta^{\dot\alpha}
\label{eqs11}
\end{eqnarray}
where
$$
\delta^{+0}_k = (\delta^{0+}_k)^*, \qquad
\delta^{-0}_k = (\delta^{0-}_k)^*
$$
The solution appears to be:
\begin{equation}
\delta^{+0}_k = \epsilon \tilde{C}, \qquad
\delta^{-0}_k = \frac{1}{(k+1)}[\tilde{m}\pm(k+1)\lambda] \tilde{C}, 
\qquad \tilde{C}^* = \mp \epsilon \tilde{C} \label{eqs12}
\end{equation}
where the $\pm$-sign corresponds to that of the relation 
$m_0{}^2=\tilde{m}(\tilde{m}\pm\lambda)$ and $\epsilon$ comes from the
$\tilde{\beta}^{+-}$ function. \\
{\bf Supertransformations for scalar} Similarly, for the spin-0 field
we take the following supertransformations (also with the complex
coefficients):
\begin{equation}
\delta W^{\alpha(k)\dot\alpha(k)} = 
\delta^{+0}_k \phi^{\alpha(k)\beta\dot\alpha(k)} \zeta_\beta
+ \delta^{-0}_k \phi^{\alpha(k-1)\dot\alpha(k)} \zeta^\alpha
+ \delta^{-+}_k \phi^{\alpha(k)\dot\alpha(k)\dot\beta} \
\zeta_{\dot\beta} + \delta^{0-}_k \phi^{\alpha(k)\dot\alpha(k-1)}
\zeta^{\dot\alpha}
 \label{eqp1}
\end{equation}
where
$$
\delta^{0+}_k = - (\delta^{+0}_k)^*, \qquad
\delta^{0-}_k = - (\delta^{-0}_k)^*
$$
with the solution:
\begin{equation}
\delta^{+0}_k = C, \qquad
\delta^{-0}_k = - \frac{\epsilon}{(k+1)}[\tilde{m}\pm(k+1)\lambda]C,
\qquad C^* = \pm \epsilon C \label{eqp2}
\end{equation}

Now having the explicit form of the supertransformations at our
disposal, it is easy to calculate their commutators and find that
their superalgebra is not closed. The reason is clear: we must have an
equal number of bosonic and fermionic degrees of freedom in each
supermultiplet. As is well known the scalar supermultiplet contains
two scalar fields, moreover, it is important that they must be scalar
and pseudo-scalar. So we consider the supermultiplet $(1/2,0,0')$. For
concreteness we take $\epsilon = +1$, then to have opposite parities
for two scalars we choose:
\begin{equation}
m_1{}^2 = \tilde{m}(\tilde{m} + \lambda), \qquad
m_2{}^2 = \tilde{m}(\tilde{m} - \lambda)
\end{equation}
The complete set of the supertransformations for the spinor now has
the form:
\begin{eqnarray}
\delta Y^{\alpha(k+1)\dot\alpha(k)} &=& 
 i\tilde{\delta}_{1,k}^- W_1^{\alpha(k)\dot\alpha(k)} \zeta^\alpha
- i\tilde{C}_1 W_1^{\alpha(k+1)\dot\alpha(k)\dot\beta} 
\zeta_{\dot\beta} \nonumber
\\
 && + \tilde{\delta}_{2,k}^- W_2^{\alpha(k)\dot\alpha(k)} \zeta^\alpha
+ \tilde{C}_2 W_2^{\alpha(k+1)\dot\alpha(k)\dot\beta} 
\zeta_{\dot\beta} \nonumber 
\\
\delta Y^{\alpha(k)\dot\alpha(k+1)} &=& 
i\tilde{C}_1 W_1^{\alpha(k)\beta\dot\alpha(k+1)} \zeta_\beta 
- i\tilde{\delta}_{1,k}^- W_1^{\alpha(k)\dot\alpha(k)} 
\zeta^{\dot\alpha} 
\\
 && + \tilde{C}_2 W_2^{\alpha(k)\beta\dot\alpha(k+1)} \zeta_\beta 
+ \tilde{\delta}_{2,k}^- W_2^{\alpha(k)\dot\alpha(k)} 
\zeta^{\dot\alpha} \nonumber
\end{eqnarray}
where
\begin{eqnarray}
\tilde{\delta}_{1,k}^- &=& \frac{1}{(k+1)} [\tilde{m} - (k+1)\lambda]
\tilde{C}_1 \nonumber
 \\
\tilde{\delta}_{2,k}^- &=& \frac{1}{(k+1)} [\tilde{m} + (k+1)\lambda]
\tilde{C}_2
\end{eqnarray}
For the supertransformations of the two scalars we have:
\begin{eqnarray}
\delta W_1^{\alpha(k)\dot\alpha(k)} &=& 
C_1 Y^{\alpha(k)\beta\dot\alpha(k)} \zeta_\beta
+ \delta_{1,k}^- Y^{\alpha(k-1)\dot\alpha(k)} \zeta^\alpha \nonumber
\\
 && - C_1 Y^{\alpha(k)\dot\alpha(k)\dot\beta} \zeta_{\dot\beta}
- \delta_{1,k}^- Y^{\alpha(k)\dot\alpha(k-1)} \zeta^{\dot\alpha}
\nonumber
 \\
\delta W_2^{\alpha(k)\dot\alpha(k)} &=& 
iC_2 Y^{\alpha(k)\beta\dot\alpha(k)} \zeta_\beta
+ i\delta_{2,k}^- Y^{\alpha(k-1)\dot\alpha(k)} \zeta^\alpha
\\
 && + iC_2 Y^{\alpha(k)\dot\alpha(k)\dot\beta} \zeta_{\dot\beta}
+ i\delta_{2,k}^- Y^{\alpha(k)\dot\alpha(k-1)} \zeta^{\dot\alpha}
\nonumber
\end{eqnarray}
where
\begin{eqnarray}
\delta_{1,k}^- &=& - \frac{1}{(k+1)} [\tilde{m} + (k+1)\lambda] C_1
\nonumber
 \\
\delta_{2,k}^- &=& - \frac{1}{(k+1)} [\tilde{m} - (k+1)\lambda] C_2
\end{eqnarray}
So we have four (real) arbitrary parameters $C_{1,2}$ and 
$\tilde{C}_{1,2}$. We proceed with calculations of the commutators.
For the first scalar field we find:
\begin{eqnarray}
\ [\delta_1, \delta_2] W_1^{\alpha(k)\dot\alpha(k)} &=&
- 2i C_1\tilde{C}_1 [ \xi_{\beta\dot\beta}
W_1^{\alpha(k)\beta\dot\alpha(k)\dot\beta} +
\beta^{--}_k \xi^{\alpha\dot\alpha} 
W_1^{\alpha(k-1)\dot\alpha(k-1)} \nonumber
 \\
 && \qquad \qquad + \lambda (\eta^\alpha{}_\beta
W_1^{\alpha(k-1)\beta\dot\alpha(k)} +
\eta^{\dot\alpha}{}_{\dot\beta} 
W_1^{\alpha(k)\dot\alpha(k-1)\dot\beta}) ]
\end{eqnarray}
where
\begin{equation}
\xi^{\alpha\dot\alpha} = \zeta_1^\alpha \zeta_2^{\dot\alpha}
- (1 \leftrightarrow 2), \qquad \eta^{\alpha(2)} = \zeta_1^\alpha
\zeta_2^\alpha - (1 \leftrightarrow 2)
\end{equation}
The results for the second scalar $W_2$ are the same provided
\begin{equation}
C_2\tilde{C}_2 = - C_1\tilde{C}_1
\end{equation}
At last, for the spinor we obtain:
\begin{eqnarray}
\ [\delta_1, \delta_2] Y^{\alpha(k+1)\dot\alpha(k)} &=& 
- 2iC_1\tilde{C}_1 [ \xi_{\beta\dot\beta}
Y^{\alpha(k+1)\beta\dot\alpha(k)\dot\beta} + 
\tilde{\beta}^{-+}_k \xi^\alpha{}_{\dot\beta}
Y^{\alpha(k)\dot\alpha(k)\dot\beta} + \tilde{\beta}^{--}_k
\xi^{\alpha\dot\alpha} Y^{\alpha(k)\dot\alpha(k-1)} \nonumber
 \\
 && \qquad \qquad + \lambda (\eta^\alpha{}_\beta
Y^{\alpha(k)\beta\dot\alpha(k)} + 
\eta^{\dot\alpha}{}_{\dot\beta}
Y^{\alpha(k+1)\dot\alpha(k-1)\dot\beta}) ] 
\end{eqnarray}
Comparison with the initial unfolded equations shows that the
supertransformations close on-shell and give $AdS_4$ superalgebra:
$$
\{ Q^\alpha, Q^{\dot\alpha} \} \sim P^{\alpha\dot\alpha}, \qquad
\{ Q^\alpha, Q^\beta \} \sim \lambda M^{\alpha\beta}, \qquad
\{ Q^{\dot\alpha}, Q^{\dot\beta} \} \sim \lambda 
M^{\dot\alpha\dot\beta}
$$

\subsection{Vector supermultiplet}

Let us turn to our second example --- vector supermultiplet. We begin
with the pair vector-spinor. \\
{\bf Supertransformations for vector} The most general ansatz (taking
into account the hermicity conditions) has the form:
\begin{eqnarray}
\delta W^{\alpha(k+1)\dot\alpha(k-1)} &=& \delta^{-0}_{k,1}
Y^{\alpha(k)\dot\alpha(k-1)} \zeta^\alpha - (\delta_{k,1}^{+0})^*
Y^{\alpha(k+1)\dot\alpha(k-1)\dot\beta} \zeta_{\dot\beta} \nonumber
\\
\delta W^{\alpha(k)\dot\alpha(k)} &=& \delta^{+0}_{k,0}
Y^{\alpha(k)\beta\dot\alpha(k)} \zeta_\beta + \delta^{-0}_{k,0}
Y^{\alpha(k-1)\dot\alpha(k)} \zeta^\alpha \nonumber
\\
 && - (\delta^{+0}_{k,0})^* Y^{\alpha(k)\dot\alpha(k)\dot\beta} 
\zeta_{\dot\beta} - (\delta^{-0}_{k,0})^* Y^{\alpha(k)\dot\alpha(k-1)}
\zeta^{\dot\alpha} \label{eqv1}
\\
\delta W^{\alpha(k-1)\dot\alpha(k+1)} &=& \delta^{+0}_{k,1}
Y^{\alpha(k-1)\beta\dot\alpha(k+1)} \zeta_\beta -
(\delta^{-0}_{k,1})^* Y^{\alpha(k-1)\dot\alpha(k)} \zeta^{\dot\alpha}
\nonumber
\end{eqnarray}
where all the coefficients are in general complex. The invariance of
the unfolded equations gives:
\begin{eqnarray}
\delta^{+0}_{k,1} &=& 2\epsilon(\tilde{m}\pm\lambda)C,
\qquad \delta^{+0}_{k,0} = C, \qquad C^* = \mp \epsilon C \nonumber
\\
\delta^{-0}_{k,1} &=& \frac{2}{(k+1)}[\tilde{m}\pm\lambda]
[\tilde{m}\pm(k+1)\lambda] C  \label{eqv2} 
\\
\delta^{-0}_{k,0} &=& \epsilon \frac{(k+2)}{k(k+1)}
[\tilde{m}\pm(k+1)\lambda] C
\nonumber
\end{eqnarray}
{\bf Supertransformations for spinor} Similarly, we introduce:
\begin{eqnarray}
\delta Y^{\alpha(k+1)\dot\alpha(k)} &=& \tilde{\delta}^{+0}_{k,1}
W^{\alpha(k+1)\beta\dot\alpha(k)} \zeta_\beta + 
\tilde{\delta}^{-0}_{k,1} W^{\alpha(k)\dot\alpha(k)} \zeta^\alpha
\nonumber
\\
 && + (\tilde{\delta}^{+0}_{k,0})^* 
W^{\alpha(k+1)\dot\alpha(k)\dot\beta} \zeta_{\dot\beta} + 
(\tilde{\delta}^{-0}_{k,0})^* W^{\alpha(k+1)\dot\alpha(k-1)}
\zeta^{\dot\alpha}
\nonumber \\
\delta Y^{\alpha(k)\dot\alpha(k+1)} &=& \tilde{\delta}^{+0}_{k,0}
W^{\alpha(k)\beta\dot\alpha(k+1)} \zeta_\beta + 
\tilde{\delta}^{-0}_{k,0} W^{\alpha(k-1)\dot\alpha(k+1)} \zeta^\alpha
\label{eqs21}
\\
 && + (\tilde{\delta}^{+0}_{k,1})^* 
W^{\alpha(k)\dot\alpha(k+1)\dot\beta} \zeta_{\dot\beta} +
(\tilde{\delta}^{-0}_{k,1})^* W^{\alpha(k)\dot\alpha(k)}
\zeta^{\dot\alpha} \nonumber
\end{eqnarray}
and obtain:
\begin{eqnarray}
\tilde{\delta}^{+0}_{k,1} &=& \tilde{C}, \qquad 
\tilde{\delta}^{+0}_{k,0} = \epsilon m_1 \tilde{C}, \qquad 
\tilde{C}^* = \pm \epsilon \tilde{C} \nonumber
\\
\tilde{\delta}^{-0}_{k,1} &=& - \frac{k}{(k+1)(k+2)}
\tilde{m}[\tilde{m}\mp(k+1)\lambda]\tilde{C} \label{eqs22}
\\
\tilde{\delta}^{-0}_{k,0} &=& - \epsilon \frac{1}{(k+1)}
[\tilde{m}\mp(k+1)\lambda]\tilde{C} \nonumber
\end{eqnarray}

It is straightforward to check that these supertransformations do not
close and the reason is again that we have three physical degrees of
freedom for the massive vector and only two --- for spinor. So we turn
to the complete vector supermultiplet $(1,1/2,1/2,0')$. In this case
it is also important that the spin 1 and spin 0 have opposite
parities. We assume that the coefficients for the vector field
supertransformations are real and choose:
\begin{equation}
m_v{}^2 = m_1(m_1+\lambda) = m_2(m_2-\lambda) = m_s{}^2, \qquad
\epsilon_1 = - 1, \qquad \epsilon_2 = + 1
\end{equation}
where $m_{1,2}$ are masses of the two spinors. This leads to the
following expressions for the four possible boson-fermion pairs. For
the vector and first spinor we have formulas (\ref{eqv1}),(\ref{eqv2})
with the parameter $C_1$ and (\ref{eqs21}),(\ref{eqs22}) with the
parameter $i\tilde{C}_1$ (all with upper signs), while for the second
spinor --- the same formulas but with the parameters $C_2$, 
$i\tilde{C}_2$ (with lower signs). Similarly, for the first spinor and
the pseudo-scalar we have formulas (\ref{eqs11}),(\ref{eqs12}) with
the parameter $iC_3$ and (\ref{eqp1}),(\ref{eqp2}) with the parameter
$\tilde{C}_3$ (with upper signs), while for the second spinor --- the
same with the parameters $iC_4$, $\tilde{C}_4$ (with lower signs).

So we have eight (real) parameters $C_{1-4}$, $\tilde{C}_{1-4}$. Let
us consider the commutators for these supertransformations. Note that
all subsequent formulas are given up to the common multiplier
$-2i(m_1+m_2)C_1\tilde{C}_1$. 

The closure of the superalgebra on the vector field requires:
$$
C_1\tilde{C}_1 + C_2\tilde{C}_2 = 0, \qquad
m_2C_1\tilde{C}_3 + m_1C_2\tilde{C}_4 = 0
$$
In this case we obtain:
\begin{eqnarray}
\ [\delta_1, \delta_2] W^{\alpha(k+1)\dot\alpha(k-1)} &\sim&
\xi_{\beta\dot\beta} W^{\alpha(k+1)\beta\dot\alpha(k-1)\dot\beta} +
\beta^{-+}_{k,1} \xi^\alpha{}_{\dot\beta}
W^{\alpha(k)\dot\alpha(k-1)\dot\beta} + \beta^{--}_{k,1}
\xi^{\alpha\dot\alpha} W^{\alpha(k)\dot\alpha(k-2)} \nonumber \\
 && + \lambda [ \eta^\alpha{}_\beta W^{\alpha(k)\beta\dot\alpha(k-1)}
+ \eta^{\dot\alpha}{}_{\dot\beta} 
W^{\alpha(k+1)\dot\alpha(k-2)\dot\beta}] \\
\ [\delta_1, \delta_2] W^{\alpha(k)\dot\alpha(k)} &\sim& 
\xi_{\beta\dot\beta} W^{\alpha(k)\beta\dot\alpha(k)\dot\beta} +
\beta^{-+}_{k,0} \xi^\alpha{}_{\dot\beta} 
W^{\alpha(k-1)\dot\alpha(k)\dot\beta} \nonumber
\\
 && + \beta^{+-}_{k,0} \xi_\beta{}^{\dot\alpha}
W^{\alpha(k)\beta\dot\alpha(k-1)} + \beta^{--}_{k,0} 
\xi^{\alpha\dot\alpha} W^{\alpha(k-1)\dot\alpha(k-1)} 
\nonumber \\
 && + \lambda [\eta^\alpha{}_\beta 
W^{\alpha(k-1)\beta\dot\alpha(k-1)} +
\eta^{\dot\alpha}{}_{\dot\beta} 
W^{\alpha(k)\dot\alpha(k-1)\dot\beta} ] 
\end{eqnarray}
For the first spinor we obtain the conditions
$$
m_1C_1\tilde{C}_1 + C_3\tilde{C}_3 = 0, \qquad 
m_1C_2\tilde{C}_1 + C_4\tilde{C}_3 = 0
$$
leading to
\begin{eqnarray}
\ [\delta_1, \delta_2] Y^{\alpha(k+1)\dot\alpha(k)} &\sim& 
\xi_{\beta\dot\beta} Y^{\alpha(k+1)\beta\dot\alpha(k)\dot\beta} +
\gamma^{-+}_k \xi^\alpha{}_{\dot\beta} 
Y^{\alpha(k)\dot\alpha(k)\dot\beta} + \gamma^{--}_k 
\xi^{\alpha\dot\alpha} Y^{\alpha(k)\dot\alpha(k-1)} \nonumber
\\
 && + \lambda [\eta^\alpha{}_\beta Y^{\alpha(k)\beta\dot\alpha(k)}
+ \eta^{\dot\alpha}{}_{\dot\beta}
Y^{\alpha(k+1)\dot\alpha(k-1)\dot\beta} ]
\end{eqnarray}
The results for the second spinor are the same provided
$$
m_2C_2\tilde{C}_2 + C_4\tilde{C}_4 = 0, \qquad 
m_2C_1\tilde{C}_2 + C_3\tilde{C}_4 = 0
$$
At last the commutator on the pseudo-scalar closes if
$$
C_3\tilde{C}_1 + C_4\tilde{C}_2 = 0
$$
and gives
\begin{eqnarray}
\ [\delta_1, \delta_2] \tilde{W}^{\alpha(k)\dot\alpha(k)} &\sim& 
\xi_{\beta\dot\beta} \tilde{W}^{\alpha(k)\beta\dot\alpha(k)\dot\beta}
+ \beta^{--}_k \xi^{\alpha\dot\alpha}
\tilde{W}^{\alpha(k-1)\dot\alpha(k-1)} \nonumber 
\\
 && + \lambda [\eta^\alpha{}_\beta 
\tilde{W}^{\alpha(k-1)\beta\dot\alpha(k)}
+ \eta^{\dot\alpha}{}_{\dot\beta} 
\tilde{W}^{\alpha(k)\dot\alpha(k-1)\dot\beta} ]
\end{eqnarray}
Thus we indeed obtain the correct on-shell superalgebra provided a
number of relations on the parameters hold. It is easy to check that
these relations are consistent, one of the possible simple solutions
being
$$
C_2 = C_3 = C_4 = C_1, \qquad
\tilde{C}_2 = - \tilde{C}_1, \qquad 
\tilde{C}_3 = - m_1\tilde{C}_1, \qquad 
\tilde{C}_4 = m_2\tilde{C}_1
$$

\section{Massive higher spin supermultiplets}

Lagrangian formulation for the massive higher spin $N=1$
supermultiplets in $AdS_4$ has been developed in \cite{BKhSZ19}. In
this section we consider an unfolded formulation for these
supermultiplets. First of all we recall the unfolded equations for
massive bosonic and fermionic fields developed in \cite{KhZ19}. Then
we consider the pairs of bosonic and fermionic fields which differ in
spin by $\iz$ (we call them superblock) and construct the
supertransformations transforming bosonic equations into fermionic
ones and vice versa. At last we consider two types of massive
supermultiplets (with integer and half-integer superspins) and adjust
the parameters of their four superblocks so that the superalgebra is
closed.

\subsection{Unfolded equations}

Let us recall the unfolded equations developed in \cite{KhZ19}.

\subsubsection{Bosonic case}

To describe a massive spin-$s$ boson, one needs gauge one-forms
(physical, auxiliary and extra) 
$\Omega^{\alpha(k+m)\dot\alpha(k-m)}$, $|m|\le k \le s-1$, 
Stueckelberg zero-forms $W^{\alpha(k+m)\dot\alpha(k-m)}$, 
$|m|\le k \le s-1$, and gauge invariant zero-forms 
$W^{\alpha(k+m)\dot\alpha(k-m)}$, $|m|\le s \le k$. We use a
convenient normalization of the Stueckelberg zero-forms where their
transformations are just shifts:
\begin{equation}
\delta W^{\alpha(k+m)\dot\alpha(k-m)} =
\eta^{\alpha(k+m)\dot\alpha(k-m)}
\end{equation}
As for the gauge one-forms, their gauge transformations:
\begin{eqnarray}
\label{gauge_trans1}
\delta\Omega^{\alpha(k+m){\dot{\alpha}}(k-m)} &=&
D\eta^{\alpha(k+m)\dot\alpha(k-m)}
+ \alpha^{--}_{k,m} e^{\alpha\dot\alpha}
\eta^{\alpha(k+m-1)\dot\alpha(k-m-1)} \nonumber
\\
 && + \alpha^{++}_{k,m} e_{\beta\dot\beta}
\eta^{\alpha(k+m)\beta\dot\alpha(k-m)\dot\beta}
+ \alpha^{-+}_{k,m} e^\alpha{}_{\dot\beta}
\eta^{\alpha(k+m-1)\dot\alpha(k-m)\dot\beta} \nonumber
\\
 && + \alpha^{+-}_{k,m} e_\beta{}^{\dot\alpha}
\eta^{\alpha(k+m)\beta\dot\alpha(k-m-1)},
\\
\delta\Omega^{\alpha(2k)} &=&
D\eta^{\alpha(2k)} + \alpha^{++}_{k,k} e_{\beta\dot\alpha}
\eta^{\alpha(2k)\beta\dot\alpha} + \alpha^{-+}_{k,k} 
e^\alpha{}_{\dot\alpha} \eta^{\alpha(2k-1)\dot\alpha}, \nonumber
\end{eqnarray}
are the modification of the massless ones by the cross terms with the
coefficients $\alpha^{+-}$, $\alpha^{-+}$. In what follows we  assume
that all functions $\alpha$ are real and satisfy the hermiticity
conditions:
$$
\alpha^{+-}_{k,m} = \alpha^{-+}_{k,-m}, \qquad
\alpha^{++}_{k,m} = \alpha^{++}_{k,-m}, \qquad
\alpha^{--}_{k,m} = \alpha^{--}_{k,-m}
$$
All these functions can be expressed in terms of the main one 
$\alpha^{-+}_m$:
\begin{eqnarray}
\alpha^{-+}_{k,m} &=& \frac{\alpha^{-+}_{m}}
{(k-m+1)(k-m+2)(k+m)(k+m+1)}, \quad m>0,\nonumber
\\
\alpha^{++}_{k,m} &=& \frac{\alpha^{++}_{k}}{(k-m+1)(k-m+2)}, \quad 
m\ge 0, 
\nonumber\\
\alpha^{--}_{k,m} &=& \frac{\alpha^{--}_{k}}{(k+m)(k+m+1)}, \quad
m\ge 0,
 \\
\alpha^{+-}_{k,m} &=& 1, \quad m \ge 0 \nonumber 
\\
\alpha^{++}_k{}^2 &=& k(k+1)\alpha^{-+}_{k+2}, \quad
k\ge 2, \quad \alpha^{++}_{0}{}^2 = 2\alpha^{-+}_{2}, \quad
\alpha^{--}_{k}{}^2 = \frac{\alpha^{-+}_{k+1}}{k(k-1)}
\nonumber 
\end{eqnarray}
For the massive spin-$s$ boson we consider in this subsection
the function $\alpha^{-+}_{m}$ is:
\begin{equation}
\alpha^{-+}_{m} = (s-m+1)(s+m)[M^2-m(m-1)\lambda^2]
\end{equation}
As is well known, in the flat space masses of the all members of the
same supermultiplet must be equal. As it was shown in \cite{BKhSZ19},
in $AdS_4$ case bosonic $M$ and fermionic $\tilde{M}$ mass parameters
must  satisfy the relation $M^2 = \tilde{M}[\tilde{M} \pm \lambda]$.
In
this case the function $\alpha^{-+}_m$ takes the form:
\begin{equation}
\alpha^{-+}_m = (s-m+1)(s+m)[\tilde{M} \pm m\lambda]
[\tilde{M} \mp (m-1)\lambda]
\end{equation}
and this factorization appears to be crucial for the construction of
the superblocks and hence the supermultiplets.

The explicit expressions for all the $\alpha$-functions given above
were found \cite{KhZ19} in the construction of the gauge invariant
self consistent two-forms (curvatures) for each gauge one-form
$(0 \le m < k)$:
\begin{eqnarray}
\label{curvatures1}
\mathcal{R}^{\alpha(k+m)\dot\alpha(k-m)} &=&
D\Omega^{{\alpha(k+m)\dot\alpha(k-m)}}
+ \alpha^{--}_{k,m} e^{\alpha\dot\alpha}
\Omega^{\alpha(k+m-1)\dot\alpha(k-m-1)} \nonumber
\\
 && + \alpha^{++}_{k,m} e_{\beta\dot\beta}
\Omega^{\alpha(k+m)\beta\dot\alpha(k-m)\dot\beta}
+ \alpha^{-+}_{k,m} e^\alpha{}_{\dot\beta}
\Omega^{\alpha(k+m-1)\dot\alpha(k-m)\dot\beta} \nonumber
\\
 && + \alpha^{+-}_{k,m} e_\beta{}^{\dot\alpha}
\Omega^{\alpha(k+m)\beta\dot\alpha(k-m-1)}, \nonumber
\\
\mathcal{R}^{\alpha(2k)} &=& D\Omega^{\alpha(2k)}
+ \alpha^{++}_{k,k} e_{\beta\dot\alpha}
\Omega^{\alpha(2k)\beta\dot\alpha}
+ \alpha^{-+}_{k,k} e^\alpha{}_{\dot\alpha}
\Omega^{\alpha(2k-1)\dot\alpha} 
\\
 && - 2\alpha^{-+}_{k,k}\alpha^{--}_{k} E^{\alpha(2)}
W^{\alpha(2k-2)} - 2\alpha^{++}_{k,k} E_{\beta(2)} 
W^{\alpha(2k)\beta(2)} \nonumber
\\
 && - \frac{\alpha^{-+}_{k+1}}{k+1} E^\alpha{}_\beta 
W^{\alpha(2k-1)\beta} \nonumber
\\
\mathcal{R} &=& D\Omega
+ \alpha^{++}_{0,0} e_{\alpha\dot\alpha}
\Omega^{\alpha\dot\alpha}
-2\alpha^{++}_{0,0} E_{\alpha(2)} W^{\alpha(2)}
-2\alpha^{++}_{0,0} E_{\dot\alpha(2)} W^{\dot\alpha(2)}.
\nonumber
\end{eqnarray}
Due to the simple normalization for the Stueckelberg zero-forms we
use, their gauge invariant one-forms are determined by the same
$\alpha$-functions:
\begin{eqnarray}
\mathcal{C}^{\alpha(k+m)\dot\alpha(k-m)} &=&
DW^{\alpha(k+m)\dot\alpha(k-m)} - \Omega^{\alpha(k+m)\dot\alpha(k-m)}
 + \alpha^{--}_{k,m} e^{\alpha\dot\alpha}
W^{\alpha(k+m-1)\dot\alpha(k-m-1)} \nonumber
\\
 && + \alpha^{++}_{k,m} e_{\beta\dot\beta}
W^{\alpha(k+m)\beta\dot\alpha(k-m)\dot\beta}
+ \alpha^{+-}_{k,m} e_\beta{}^{\dot\alpha}
W^{\alpha(k+m)\beta\dot\alpha(k-m-1)} \nonumber
\\
 && + \alpha^{-+}_{k,m} e^\alpha{}_{\dot\beta}
W^{\alpha(k+m-1)\dot\alpha(k-m)\dot\beta}.
\label{curvatures2} 
\end{eqnarray}

Now we are ready to present a set of unfolded equations. The whole
system can be subdivided into three subsystems. The first subsystem is
just the zero curvature conditions for most of the gauge invariant
two- and one-forms (except some highest ones, see below):
\begin{eqnarray}
\label{unf_eq1}
0 &=& \mathcal{R}^{\alpha(k+m)\dot\alpha(k-m)}, \qquad \qquad k<s-1
\nonumber \\
0 &=& \mathcal{R}^{\alpha(s-1+m)\dot\alpha(s-1-m)}, \qquad |m|<s-1
 \\
0 &=& \mathcal{C}^{\alpha(k+m)\dot\alpha(k-m)}, \qquad \qquad k<s-1
\nonumber
\end{eqnarray}
The second one contains these remaining gauge invariant curvatures
and gives a connection with the sector of the gauge invariant
zero-forms:
\begin{eqnarray}
\label{unf_eq2}
0 &=& \mathcal{R}^{\alpha(2s-2)}+ 2E_{\beta(2)}
W^{\alpha(2s-2)\beta(2)}
\nonumber \\
0 &=& \mathcal{C}^{\alpha(s-1+m)\dot\alpha(s-1-m)}- 
e_{\beta\dot\beta} W^{\alpha(s-1+m)\beta\alpha(s-1-m)\dot\beta}
\end{eqnarray}
Finally, the third one contains the gauge invariant zero-forms only.
Its structure reproduces the structure of the unfolded equations for
massless components with added cross terms $(m < k)$:
\begin{eqnarray}
\label{unf_eq3}
0 &=& DW^{\alpha(k+m)\dot\alpha(k-m)}
+ \beta^{--}_{k,m} e^{\alpha\dot\alpha}
W^{\alpha(k+m-1)\dot\alpha(k-m-1)} \nonumber
\\
 && + \beta^{++}_{k,m} e_{\beta\dot\beta} 
W^{\alpha(k+m)\beta\dot\alpha(k-m)\dot\beta}
+ \beta^{+-}_{k,m} e_\beta{}^{\dot\alpha}
W^{\alpha(k+m)\beta\dot\alpha(k-m-1)} \nonumber
\\
 && + \beta^{-+}_{k,m} e^\alpha{}_{\dot\beta}
W^{\alpha(k+m-1)\dot\alpha(k-m)\dot\beta}, \\
0 &=& D W^{\alpha(2k)} + \beta^{++}_{k,k} e_{\beta\dot\alpha}
W^{\alpha(2k)\beta\dot\alpha} + \beta^{-+}_{k,k}
e^\alpha{}_{\dot\alpha} W^{\alpha(2k-1)\dot\alpha} \nonumber
\end{eqnarray}
Here we also assume that all the functions $\beta$ are real and
satisfy the hermiticity conditions:
$$
\beta^{+-}_{k,m} = \beta^{-+}_{k,-m}, \qquad
\beta^{++}_{k,m} = \beta^{++}_{k,-m}, \qquad
\beta^{--}_{k,m} = \beta^{--}_{k,-m}.
$$
The coefficients $\beta^{ij}_{k,m}$ are determined by the
self-consistency of these equations (taking into account their
connection with the gauge sector). It appears that all of them can be
expressed via the very same function $\alpha^{-+}_{m}$:
\begin{eqnarray}
\beta^{-+}_{k,m} &=& \frac{\beta^{-+}_{m}}{(k+m)(k+m+1)}, \quad
m\ge 0, \nonumber
\\
\beta^{+-}_{k,m} &=& \frac{\beta^{+-}_{m}}{(k-m)(k-m+1)}, \quad 
m\ge 0, \nonumber
\\
\beta^{--}_{k,m} &=& \frac{\alpha^{-+}_{k+1}}
{(k+m)(k+m+1)(k-m)(k-m+1)}, \quad k>s, \quad \beta^{--}_{s,m}=0,
\nonumber\\
\beta^{-+}_{m} &=& \frac{\alpha^{-+}_{m}}{(s-m)(s-m+1)}, \quad 
1\le m<s, \qquad \beta_{s}^{-+} = \frac{\alpha^{-+}_s}{2}, \nonumber
\\
\beta^{+-}_{m} &=& (s-m-1)(s-m), \quad 0\le m<s-1, \qquad
\beta^{+-}_{s-1} = 2, \nonumber
\end{eqnarray}

\subsubsection{Fermionic case}

Similarly to the massive boson, to describe a massive
spin-$\tilde{s}=s+\iz$ fermion, one needs one-forms (physical and
extra ones) $\Psi^{\alpha(k+m)\dot\alpha(k-m)}$, 
$|m|\le k \le \tilde{s}-1$, Stueckelberg zero-forms 
$Y^{\alpha(k+m)\dot\alpha(k-m)}$, $|m|\le k \le \tilde{s}-1$,
and gauge invariant zero-forms $Y^{\alpha(k+m)\dot\alpha(k-m)}$, 
$|m|\le \tilde{s} \le k$; the indices $k,m$ are half-integers now. The
ansatz for gauge transformations and gauge invariant curvatures for
the fermions has the same form as the corresponding expressions for
bosons; but the coefficients $\tilde{\alpha}^{ij}_{k,m}$ are different
from the corresponding bosonic ones. The gauge transformations are:
\begin{eqnarray}
\label{gauge_trans1a}
\delta\Psi^{\alpha(k+m){\dot{\alpha}}(k-m)} &=&
D\eta^{\alpha(k+m)\dot\alpha(k-m)}
+ \tilde{\alpha}^{--}_{k,m} e^{\alpha\dot\alpha}
\eta^{\alpha(k+m-1)\dot\alpha(k-m-1)} \nonumber
\\
 && + \tilde{\alpha}^{++}_{k,m} e_{\beta\dot\beta}
\eta^{\alpha(k+m)\beta\dot\alpha(k-m)\dot\beta}
+ \tilde{\alpha}^{-+}_{k,m} e^\alpha{}_{\dot\beta}
\eta^{\alpha(k+m-1)\dot\alpha(k-m)\dot\beta} \nonumber
\\
 && + \tilde{\alpha}^{+-}_{k,m}e_\beta{}^{\dot\alpha}
\eta^{\alpha(k+m)\beta\dot\alpha(k-m-1)},
\\
\delta\Psi^{\alpha(2k)} &=&
D\eta^{\alpha(2k)} 
+ \tilde{\alpha}^{++}_{k,k}
e_{\beta\dot\alpha} \eta^{\alpha(2k)\beta\dot\alpha} 
+ \tilde{\alpha}^{-+}_{k,k} e^\alpha{}_{\dot\alpha}
\eta^{\alpha(2k-1)\dot\alpha}, \nonumber
\\
\delta Y^{\alpha(k+m)\dot\alpha(k-m)} &=&
\eta^{\alpha(k+m)\dot\alpha(k-m)}, \nonumber
\end{eqnarray}
where all the functions $\tilde{\alpha}$ are assumed to be real and
satisfying the hermiticity conditions:
$$
\tilde{\alpha}^{+-}_{k,m} = \alpha^{-+}_{k,-m}, \qquad
\tilde{\alpha}^{++}_{k,m} = \alpha^{++}_{k,-m}, \qquad
\tilde{\alpha}^{--}_{k,m} = \alpha^{--}_{k,-m}
$$
All of them also can be expressed in terms of one main function
$\tilde{\alpha}^{-+}_m$:
\begin{eqnarray}
\tilde{\alpha}^{-+}_{k,m} &=& \frac{\tilde{\alpha}^{-+}_{m}}
{(k-m+1)(k-m+2)(k+m)(k+m+1)}, \quad m>\iz, \nonumber
\\
\tilde{\alpha}^{-+}_{k,\iz} &=&
\frac{\epsilon\sqrt{\tilde{\alpha}^{-+}_{\iz}} }
{(k+\iz)(k+\tz)}, \nonumber
\\
\tilde{\alpha}^{+-}_{k,m} &=& 1, \quad m\ge \iz, \nonumber
\\
\tilde{\alpha}^{++}_{k,m} &=&
\frac{\tilde{\alpha}^{++}_{k}}{(k-m+1)(k-m+2)}, \quad m\ge \iz,
\\
\tilde{\alpha}^{--}_{k,m} &=&
\frac{\tilde{\alpha}^{--}_{k}}{(k+m)(k+m+1)}, \quad m\ge \iz,
\nonumber
\\
\tilde{\alpha}^{++}_k{}^2 &=& (k+\iz)^2\tilde{\alpha}^{-+}_{k+2},
\qquad \tilde{\alpha}^{--}_{k}{}^2 =
\frac{\tilde{\alpha}^{-+}_{k+1}}{(k-\iz)^2},
\nonumber 
\end{eqnarray}
Here, the function $\tilde{\alpha}^{-+}_{m}$ is:
\begin{eqnarray}
\tilde{\alpha}^{-+}_{m}&=(\tilde{s}-m+1)(\tilde{s}+m)\left(\tilde{M}^2-(m-\iz)^2\lambda^2\right)
\end{eqnarray}
In particular,
$\sqrt{\tilde{\alpha}^{-+}_{\iz}}=(\tilde{s}+\iz)\tilde{M}$. One of
the essential differences between bosons and fermions is that bosons
have the mass-like terms proportional to $M^2$, while fermions --- to 
$\tilde{M}$. And as it was shown in \cite{BKhSZ19}, the sign of the
fermionic mass term plays an important role in the construction of the
supermultiplets. Namely, the signs for the two fermions entering the
supermultiplet must be opposite. Thus in the expressions given above
we introduced the parameter $\epsilon=\pm 1$ corresponding to the
choice of mass-like terms sign, while we always assume that the
parameters $M$ and $\tilde{M}$ are positive.

As in the bosonic case, for each gauge one-form one can construct a
gauge invariant two-form --- curvature $(0 \le m < k)$:
\begin{eqnarray}
\label{curvatures1a}
\mathcal{F}^{\alpha(k+m)\dot\alpha(k-m)} &=&
D\Psi^{{\alpha(k+m)\dot\alpha(k-m)}}
+ \tilde{\alpha}^{--}_{k,m} e^{\alpha\dot\alpha}
\Psi^{\alpha(k+m-1)\dot\alpha(k-m-1)} \nonumber
\\
 && + \tilde{\alpha}^{++}_{k,m} e_{\beta\dot\beta}
\Psi^{\alpha(k+m)\beta\dot\alpha(k-m)\dot\beta}
+ \tilde{\alpha}^{-+}_{k,m} e^\alpha{}_{\dot\beta}
\Psi^{\alpha(k+m-1)\dot\alpha(k-m)\dot\beta} \nonumber
\\
 && + \tilde{\alpha}^{+-}_{k,m} e_\beta{}^{\dot\alpha}
\Psi^{\alpha(k+m)\beta\dot\alpha(k-m-1)}, 
\\
\mathcal{F}^{\alpha(2k)} &=& D\Psi^{\alpha(2k)}
+ \tilde{\alpha}^{++}_{k,m} e_{\beta\dot\alpha}
\Psi^{\alpha(2k)\beta\dot\alpha}
+ \tilde{\alpha}^{-+}_{k,k} e^\alpha{}_{\dot\alpha}
\Psi^{\alpha(2k-1)\dot\alpha} \nonumber
\\
 && - 2\tilde{\alpha}^{-+}_{k,k}\tilde{\alpha}^{--}_{k,k-1}
E^{\alpha(2)} Y^{\alpha(2k-2)} - 2\tilde{\alpha}^{++}_{k,k}
E_{\beta(2)} Y^{\alpha(2k)\beta(2)}
\nonumber
\\
 && - \frac{\tilde{\alpha}^{-+}_{k+1}}{k+1} E^\alpha{}_\beta 
Y^{\alpha(2k-1)\beta}, \nonumber
\end{eqnarray}
as well as a gauge invariant one-form for each Stueckelberg zero-form:
\begin{eqnarray}
\mathcal{D}^{\alpha(k+m)\dot\alpha(k-m)} &=&
DY^{\alpha(k+m)\dot\alpha(k-m)} - \Psi^{\alpha(k+m)\dot\alpha(k-m)}
 + \tilde{\alpha}^{--}_{k,m} e^{\alpha\dot\alpha}
Y^{\alpha(k+m-1)\dot\alpha(k-m-1)} \nonumber
\\
 && + \tilde{\alpha}^{++}_{k,m} e_{\beta\dot\beta}
Y^{\alpha(k+m)\beta\dot\alpha(k-m)\dot\beta}
+ \tilde{\alpha}^{+-}_{k,m} e_\beta{}^{\dot\alpha}
Y^{\alpha(k+m)\beta\dot\alpha(k-m-1)} \nonumber
\\
 && + \tilde{\alpha}^{-+}_{k,m} e^\alpha{}_{\dot\beta}
Y^{\alpha(k+m-1)\dot\alpha(k-m)\dot\beta}. 
\label{curvatures2a}
\end{eqnarray}

Now let us consider a set of the unfolded equation. Here the whole
system also can be subdivided into three subsystems. The first
subsystem is just the zero curvature conditions for most of the gauge
invariant two- and one-forms:
\begin{eqnarray}
\label{unf_eq1a}
0 &=& \mathcal{F}^{\alpha(k+m)\dot\alpha(k-m)}, \qquad\qquad k<s-1
\nonumber \\
0 &=& \mathcal{F}^{\alpha(s-1+m)\dot\alpha(s-1-m)}, \qquad |m|<s-1
 \\
0 &=& \mathcal{D}^{\alpha(k+m)\dot\alpha(k-m)}, \qquad\qquad k<s-1
\nonumber
\end{eqnarray}
The second one contains the remaining gauge invariant curvatures and
gives a connection with the sector of the gauge invariant zero-forms:
\begin{eqnarray}
\label{unf_eq2a}
0 &=& \mathcal{F}^{\alpha(2s-2)} + E_{\beta(2)}
Y^{\alpha(2s-2)\beta(2)}
\nonumber \\
0 &=& \mathcal{D}^{\alpha(s-1+m)\dot\alpha(s-1-m)} - 
e_{\beta\dot\beta} Y^{\alpha(s-1+m)\beta\alpha(s-1-m)\dot\beta}
\end{eqnarray}
Finally, the third one contains the gauge invariant zero-forms only.
Its structure reproduces the structure of the unfolded equations for
massless components with added cross terms $(m < k)$:
\begin{eqnarray}
\label{unf_eq3a}
0 &=& DY^{\alpha(k+m)\dot\alpha(k-m)}
+ \tilde{\beta}^{--}_{k,m} e^{\alpha\dot\alpha}
Y^{\alpha(k+m-1)\dot\alpha(k-m-1)} \nonumber
\\
 && + e_{\beta\dot\beta}
Y^{\alpha(k+m)\beta\dot\alpha(k-m)\dot\beta}
+ \tilde{\beta}^{+-}_{k,m} e_\beta{}^{\dot\alpha}
Y^{\alpha(k+m)\beta\dot\alpha(k-m-1)} \nonumber
\\
 && + \tilde{\beta}^{-+}_{k,m} e^\alpha{}_{\dot\beta}
Y^{\alpha(k+m-1)\dot\alpha(k-m)\dot\beta} \\
0 &=& D Y^{\alpha(2k)} + e_{\beta\dot\alpha} 
Y^{\alpha(2k)\beta\dot\alpha} + \tilde{\beta}^{-+}_{k,k}
e^\alpha{}_{\dot\alpha} Y^{\alpha(2k-1)\dot\alpha} \nonumber
\end{eqnarray}
 The coefficients $\tilde{\beta}^{ij}_{k,m}$ (which assumed to be real
and satisfying the hermiticity conditions similar to that of 
$\tilde{\alpha}$) are determined by the
self-consistency of these equations (taking into account the
connection with the gauge sector). They resemble the corresponding
bosonic coefficients, the most significant difference being
the behavior of some of the coefficients at $m=\pm\iz$. As in the
bosonic case, they all can be expressed via the same main function
$\tilde{\alpha}^{-+}_m$:
\begin{eqnarray}
\tilde{\beta}^{-+}_{k,m} &=&
\frac{\tilde{\beta}^{-+}_{m}}{(k+m)(k+m+1)}, \quad m\ge \iz, \nonumber
\\
\tilde{\beta}^{+-}_{k,m} &=&
\frac{\tilde{\beta}^{+-}_{m}}{(k-m)(k-m+1)}, \quad m\ge \iz, \nonumber
\\
\tilde{\beta}^{--}_{k,m} &=& \frac{\tilde{\alpha}^{-+}_{k+1}}
{(k+m)(k+m+1)(k-m)(k-m+1)},
\quad k>\tilde{s}, \quad \tilde{\beta}^{--}_{\tilde{s},m}=0,
\\
\tilde{\beta}^{-+}_{m} &=&
\frac{\tilde{\alpha}^{-+}_{m}}{(\tilde{s}-m)(\tilde{s}-m+1)}, \quad 
\iz\le m<\tilde{s}, \qquad 
\tilde{\beta}_{\iz}^{-+} = \epsilon\sqrt{\tilde{\alpha}^{-+}_{\iz}},
\quad
\tilde{\beta}_{\tilde{s}}^{-+} =
\frac{\tilde{\alpha}^{-+}_{\tilde{s}}}{2}, 
\nonumber
\\
\tilde{\beta}^{+-}_{m} &=& (\tilde{s}-m-1)(\tilde{s}-m), \quad \iz\le
m<\tilde{s}-1, \qquad
\tilde{\beta}^{+-}_{\tilde{s}-1} = 2, \nonumber
\end{eqnarray}

\subsection{Superblocks}

Similarly to the massless supermultiplets, it is possible to construct
a system of massive higher spin boson and fermion which is invariant
under the supertransformations, which we call a superblock. However, 
in contrast to the massless case, the algebra of such
supertransformations is not closed. To make it closed, one needs four
particles --- two bosons and two fermions
\cite{BKhSZ19,BKhSZ19a,BKhSZ19b}. Each pair of one boson and one
fermion forms a superblock with its own transformations, so that each
particle enters two such superblocks. Moreover, it is possible to
adjust the parameters of these superblocks so that the superalgebra is
closed.

We begin with the construction of the superblocks. Naturally,
supersymmetry requires that the parameters of the particles are
connected. First, a well-known relation $\tilde{s}=s\pm\iz$ holds for
the spins of fermion and boson. Secondly, as it was shown in
\cite{BKhSZ19}, the mass parameters of the particles are also must be
connected: $M^2=\tilde{M}(\tilde{M}\pm\lambda)$. At first, we consider
the general properties of these superblocks and then provide the
explicit solutions for the two possible types with 
$\tilde{s} = s \pm \iz$.

As we have seen, the whole set of unfolded equations both for the
bosons as well for the fermions can be subdivided into the three
sub-sectors. It is natural to begin with the subsector of the gauge
invariant zero-forms since they must form a closed subsystem under the
supertransformations as well. The most general ansatz is thus:
\begin{eqnarray}
\label{ansatz_gi0f}
\delta W^{\alpha(k+m)\dot\alpha(k-m)} &=& \delta_{k,m}^{0+}
Y^{\alpha(k+m)\dot\alpha(k-m)\dot\beta} \zeta_{\dot\beta}
+ \delta_{k,m}^{0-} Y^{\alpha(k+m)\dot\alpha(k-m-1)} 
\zeta^{\dot\alpha}
\nonumber \\
&& + \delta_{k,m}^{+0}
Y^{\alpha(k+m)\beta\dot\alpha(k-m)}\zeta_{\beta} + \delta_{k,m}^{-0}
Y^{\alpha(k+m-1)\dot\alpha(k-m)} \zeta^{\alpha}
\nonumber \\
\delta Y^{\alpha(k+m)\dot\alpha(k-m)} &=&
{\tilde{\delta}}_{k,m}^{0+} W^{\alpha(k+m)\dot\alpha(k-m)\dot\beta}
\zeta_{\dot\beta} + {\tilde{\delta}}_{k,m}^{0-}
W^{\alpha(k+m)\dot\alpha(k-m-1)}\zeta^{\dot\alpha}
 \\
&& + {\tilde{\delta}}_{k,m}^{+0}
W^{\alpha(k+m)\beta\dot\alpha(k-m)}\zeta_{\beta} +
{\tilde{\delta}}_{k,m}^{-0}
W^{\alpha(k+m-1)\dot\alpha(k-m)} \zeta^{\alpha} \nonumber
\end{eqnarray}
Here $k,m$ are integers in the first equation and half-integers in the
second one. All these functions $\delta$, $\tilde{\delta}$ are in
general complex and satisfy the hermiticity conditions:
\begin{equation}
\delta^{0+}_{k,-m} = - (\delta^{+0}_{k,m})^*, \qquad
\delta^{0-}_{k,-m} = - (\delta^{-0}_{k,m})^*, \qquad
\tilde{\delta}^{0+}_{k,-m} = (\tilde{\delta}^{+0}_{k,m})^*, \qquad
\tilde{\delta}^{0-}_{k,-m} = (\tilde{\delta}^{-0}_{k,m})^*.
\end{equation}
For lower $k$, some of the fields $W$ or $Y$ on the right-hand side
may turn out to be the Stueckelberg ones. Such terms are forbidden by
gauge invariance, so we must impose the following boundary conditions
depending on the type of the superblock: 
\begin{eqnarray}
\tilde{\delta}^{-0}_{\tilde{s},m} = 0 && \tilde{s} = s - \iz \nonumber
\\
\delta^{-0}_{s,m} = 0 && \tilde{s} = s + \iz
\end{eqnarray}
The requirement that the gauge invariant subsector of the unfolded
equations is preserved by these supertransformations leads to the
number of equations on the functions $\delta$, $\tilde{\delta}$ given
in Appendix. These equations completely determine these functions up
to the two arbitrary constants. Their explicit solutions given in the
two subsequent subsubsections. Note, that the relation
$M^2 = \tilde{M}[\tilde{M} \pm \lambda]$ appears already at this
level.

Then, we consider the supertransformations for the gauge sector. The
most general ansatz for the Stueckelberg zero-forms is:
\begin{eqnarray}
\label{ansatz_g0f}
\delta W^{\alpha(k+m)\dot\alpha(k-m)} &=& \gamma_{k,m}^{0+}
Y^{\alpha(k+m)\dot\alpha(k-m)\dot\beta} \zeta_{\dot\beta}
+ \gamma_{k,m}^{0-} Y^{\alpha(k+m)\dot\alpha(k-m-1)} 
\zeta^{\dot\alpha}
\nonumber \\
&& + \gamma_{k,m}^{+0}
Y^{\alpha(k+m)\beta\dot\alpha(k-m)}\zeta_{\beta} + \gamma_{k,m}^{-0}
Y^{\alpha(k+m-1)\dot\alpha(k-m)}\zeta^{\alpha}
\nonumber \\
\delta Y^{\alpha(k+m)\dot\alpha(k-m)} &=&
{\tilde{\gamma}}_{k,m}^{0+}W^{\alpha(k+m)\dot\alpha(k-m)\dot\beta}
\zeta_{\dot\beta}
+ {\tilde{\gamma}}_{k,m}^{0-}
W^{\alpha(k+m)\dot\alpha(k-m-1)}\zeta^{\dot\alpha}
 \\
&& + {\tilde{\gamma}}_{k,m}^{+0}
W^{\alpha(k+m)\beta\dot\alpha(k-m)}\zeta_{\beta} +
{\tilde{\gamma}}_{k,m}^{-0}
W^{\alpha(k+m-1)\dot\alpha(k-m)} \zeta^{\alpha} \nonumber
\end{eqnarray}
where all functions $\gamma$, $\tilde{\gamma}$ are in general complex
and satisfy the hermiticity conditions similar to that for the 
$\delta$, $\tilde{\delta}$:
\begin{equation}
\gamma^{0+}_{k,-m} = - (\gamma^{+0}_{k,m})^*, \qquad
\gamma^{0-}_{k,-m} = - (\gamma^{-0}_{k,m})^*, \qquad
\tilde{\gamma}^{0+}_{k,-m} = (\tilde{\gamma}^{+0}_{k,m})^*, \qquad
\tilde{\gamma}^{0-}_{k,-m} = (\tilde{\gamma}^{-0}_{k,m})^*.
\end{equation}
Most of the unfolded equations for the Stueckelberg zero-forms are
just
the zero-curvature conditions. Thus the invariance of these equations
under the supertransformations is equivalent to the following
transformations for these curvatures:
\begin{eqnarray}
\delta\mathcal{C}^{\alpha(k+m)\dot\alpha(k-m)} &=&
\gamma_{k,m}^{0+}\mathcal{D}^{\alpha(k+m)\dot\alpha(k-m)\dot\beta}
\zeta_{\dot\beta}
+ \gamma_{k,m}^{0-}
\mathcal{D}^{\alpha(k+m)\dot\alpha(k-m-1)}\zeta^{\dot\alpha}
\nonumber \\
&& + \gamma_{k,m}^{+0}
\mathcal{D}^{\alpha(k+m)\beta\dot\alpha(k-m)}\zeta_{\beta} +
\gamma_{k,m}^{-0}
\mathcal{D}^{\alpha(k+m-1)\dot\alpha(k-m)} \zeta^{\alpha}
\nonumber \\
\delta\mathcal{D}^{\alpha(k+m)\dot\alpha(k-m)} &=&
{\tilde{\gamma}}_{k,m}^{0+} 
\mathcal{C}^{\alpha(k+m)\dot\alpha(k-m)\dot\beta} \zeta_{\dot\beta}
+ {\tilde{\gamma}}_{k,m}^{0-} 
\mathcal{C}^{\alpha(k+m)\dot\alpha(k-m-1)} \zeta^{\dot\alpha}
 \\
&& + {\tilde{\gamma}}_{k,m}^{+0}
\mathcal{C}^{\alpha(k+m)\beta\dot\alpha(k-m)} \zeta_{\beta}
+ {\tilde{\gamma}}_{k,m}^{-0}
\mathcal{C}^{\alpha(k+m-1)\dot\alpha(k-m)} \zeta^{\alpha} \nonumber
\end{eqnarray}
This leads to the number of equations on the functions $\gamma$, 
$\tilde{\gamma}$ also given in Appendix. Their solutions also
determine all the functions $\gamma$, $\tilde{\gamma}$ up to the two
arbitrary constants. Note, that the supertransformations for the
Stueckelberg zero-forms can (and have to) contain gauge invariant zero
forms for highest $k=\max\{s,\tilde{s}\}$ possible. The ansatz
(\ref{ansatz_g0f}) has to be modified in a different way for the two
types of the superblocks. We will present  the modified ansatz in the
following subsubsections.

At last let us turn to the gauge one-forms. Recall that the general
form for the Stueckelberg field curvatures are 
$\mathcal{C}=DW+\Omega+\ldots$, $\mathcal{D}=DY+\Psi+\ldots$. 
This fix the supertransformations for the gauge one-forms entirely.
Except for the $|m|=k$, the structure and coefficients for the
supertransformations of one-forms are the same:
\begin{eqnarray}
\label{ansatz_g1f}
\delta\Omega^{\alpha(k+m)\dot\alpha(k-m)} &=&
\gamma_{k,m}^{0+}\Psi^{\alpha(k+m)\dot\alpha(k-m)\dot\beta}
\zeta_{\dot\beta}
+ \gamma_{k,m}^{0-} \Psi^{\alpha(k+m)\dot\alpha(k-m-1)}
\zeta^{\dot\alpha}
\nonumber \\
&& + \gamma_{k,m}^{+0}
\Psi^{\alpha(k+m)\beta\dot\alpha(k-m)}\zeta_{\beta} +
\gamma_{k,m}^{-0} \Psi^{\alpha(k+m-1)\dot\alpha(k-m)}\zeta^{\alpha}
\nonumber \\
\delta\Psi^{\alpha(k+m)\dot\alpha(k-m)} &=& 
{\tilde{\gamma}}_{k,m}^{0+}
\Omega^{\alpha(k+m)\dot\alpha(k-m)\dot\beta} \zeta_{\dot\beta}
+ {\tilde{\gamma}}_{k,m}^{0-}
\Omega^{\alpha(k+m)\dot\alpha(k-m-1)}\zeta^{\dot\alpha}
 \\
&& + {\tilde{\gamma}}_{k,m}^{+0}
\Omega^{\alpha(k+m)\beta\dot\alpha(k-m)} \zeta_{\beta}
+ {\tilde{\gamma}}_{k,m}^{-0}
\Omega^{\alpha(k+m-1)\dot\alpha(k-m)}\zeta^{\alpha} \nonumber
\end{eqnarray}
The supertransformations for one-forms $\Omega^{\alpha(2k)}$,
$\Psi^{\alpha(2k+1)}$ must contain terms with zero-forms (both
Stueckelberg and the gauge invariant ones). 
Now we consider the two cases $\tilde{s}=s\pm\iz$.

\subsubsection{Superblock $\tilde{s} = s - \iz$}

We begin with the ansatz (\ref{ansatz_gi0f}) for the gauge invariant
zero-forms. The gauge invariant sector of the unfolded equations
system is preserved under the conditions given in Appendix 
(\ref{superblock_eqs1}). Those conditions require that
$M^2=\tilde{M}(\tilde{M}\pm\lambda)$; the explicit expressions for
the coefficients $\delta^{ij}_{k,m}$ are $(m \ge 0)$:
\begin{eqnarray}
\delta^{+0}_{k,m} &=& (s-m)(s-m-1)C_b, 
\nonumber \\
\delta^{0-}_{k,m} &=& \pm
\frac{(k+s+1)(\tilde{M}\pm(k+1)\lambda)}{(k-m)(k-m+1)}
\delta^{+0}_{k,m},
\nonumber \\
\delta^{0+}_{k,m} &=& \pm (s+m)(\tilde{M}\pm m\lambda)C_b,
\quad m > 0, 
\\
\delta^{0+}_{k,0} &=& \pm \epsilon s(s-1) C_b
 \nonumber \\
\delta^{-0}_{k,m} &=& \pm
\frac{(k+s+1)(\tilde{M}\pm(k+1)\lambda)}{(k+m)(k+m+1)}
\delta^{0+}_{k,m}, \nonumber 
\end{eqnarray}
while those for the functions $\tilde{\delta}^{ij}_{k,m}$ are 
$(m \ge \iz)$:
\begin{eqnarray}
\tilde{\delta}^{+0}_{k+\iz,m+\iz} &=& C_f,
\nonumber \\
\tilde{\delta}^{0-}_{k+\iz,m+\iz} &=& \mp
\frac{(k-s+1)(\tilde{M}\mp(k+1)\lambda)}{(k-m)(k-m+1)}C_f,
\nonumber \\
\tilde{\delta}^{0+}_{k+\iz,m+\iz} &=& \pm
\frac{(\tilde{M}\mp m\lambda)}{(s-m-1)}C_f,
\\
\tilde{\delta}^{-0}_{k+\iz,m+\iz} &=& \mp
\frac{(k-s+1)(\tilde{M}\mp(k+1)\lambda)}{(k+m+1)(k+m+2)}
\tilde{\delta}^{0+}_{k+\iz,m+\iz}. \nonumber
\end{eqnarray}
The sign choice corresponds to the sign in the relation
$M^2=\tilde{M}(\tilde{M}\pm\lambda)$. Note that 
$\tilde{\delta}^{-0}_{\tilde{s},m} = 0$ as it should be. Thus all the
functions $\delta$, $\tilde{\delta}$ are determined up to the two
arbitrary complex parameters $C_b$ and $C_f$. Moreover, in $AdS$ case,
i.e. when $\lambda \ne 0$, we obtain a pair of additional relations on
these constants:
\begin{equation}
C^*_b = \mp \epsilon C_b, \qquad
C^*_f = \pm \epsilon C_f
\end{equation}

Now let us turn to the gauge sector. The invariance of the
corresponding set of the unfolded equations under the
supertransformations (\ref{ansatz_g0f}) leads to a number of equations
(\ref{superblock_eqs2}) given in Appendix. These equations determine
all the functions $\gamma$ and $\tilde{\gamma}$ also up to the two
arbitrary complex constants $C$ and $\tilde{C}$. Explicit expressions
for the functions $\gamma$ look like $(m \ge 0)$:
\begin{eqnarray}
\gamma^{+0}_{k,m} &=& \mp \sqrt{k(s-k-1)(\tilde{M}\mp(k+1)\lambda)} C,
\quad k>0, 
\nonumber \\
\gamma^{+0}_{0,0} &=& \mp \sqrt{2(s-1)(\tilde{M}\mp\lambda)} C,
\nonumber \\
\gamma^{0-}_{k,m} &=&- 
\sqrt{\frac{(s+k+1)(\tilde{M}\pm(k+1)\lambda)}{k}}C,
 \\
\gamma^{0+}_{k,m} &=& \pm
\frac{(s+m)(\tilde{M}\pm m\lambda)}{(k-m+1)(k-m+2)}
\gamma^{+0}_{k,m}, \quad m > 0, 
\nonumber \\
\gamma^{-0}_{k,m} &=& \pm
\frac{(s+m)(\tilde{M}\pm m \lambda)}{(k+m)(k+m+1)}
\gamma^{0-}_{k,m}, \quad m> 0,  
\nonumber 
\end{eqnarray}
while those for the $\tilde{\gamma}$ $(m \ge \iz)$:
\begin{eqnarray}
\tilde{\gamma}^{+0}_{k+\iz,m+\iz} &=& \mp
\sqrt{(k+1)(s+k+2)(\tilde{M}\pm(k+2)\lambda)}\tilde{C}, 
\nonumber \\
\tilde{\gamma}^{0-}_{k+\iz,m+\iz} &=& -
\sqrt{\frac{(s-k-1)(\tilde{M}\mp(k+1)\lambda)}{k}} \tilde{C},
\quad k>\iz,
\nonumber \\
\tilde{\gamma}^{0-}_{\iz,\iz} &=& -
\sqrt{\frac{(s-1)(\tilde{M}\mp\lambda)}{2}} \tilde{C}, 
\\
\tilde{\gamma}^{0+}_{k+\iz,m+\iz} &=& \pm
\frac{(s-m)(\tilde{M}\mp m \lambda)}{(k-m+1)(k-m+2)}
\tilde{\gamma}^{+0}_{k+\iz,m+\iz},
\nonumber \\
\tilde{\gamma}^{-0}_{k+\iz,m+\iz} &=& \pm
\frac{(s-m)(\tilde{M}\mp m\lambda)}{(k+m+1)(k+m+2)}
\tilde{\gamma}^{0-}_{k+\iz,m+\iz}
\nonumber
\end{eqnarray}
Similarly to the previous case, for $\lambda \ne 0$ we obtain a pair
of additional relations on these constants:
\begin{equation}
C^* = \mp \epsilon C, \qquad
\tilde{C}^* = \pm \epsilon \tilde{C}
\end{equation}
Similarly to the case with the gauge invariant two-forms, the
supertransformations for one-forms at $m = \pm k$ differ from the
general case and have to contain zero-forms:
\begin{eqnarray}
\label{superblock_diagonal}
\delta\Omega^{\alpha(2k)} &=& \gamma^{0+}_{k,k}
\Psi^{\alpha(2k)\dot\beta} \zeta_{\dot\beta}
+ \gamma^{+0}_{k,k} \Psi^{\alpha(2k)\beta} \zeta_{\beta}
+ \gamma^{-0}_{k,k} \Psi^{\alpha(2k-1)} \zeta^{\alpha}
\nonumber \\
&& + \gamma^{0-}_{k,k}
\frac{\tilde{\alpha}^{-+}_{k+\iz}}{(2k+1)}e^{\alpha}{}_{\dot\alpha}
Y^{\alpha(2k-1)} \zeta^{\dot\alpha}
+ \gamma^{0-}_{k,k}
\tilde{\alpha}^{++}_{k-\iz}e_{\beta\dot\alpha}Y^{\alpha(2k)\beta}\zeta^{\dot\alpha}, \quad k>0,
\nonumber \\
\delta\Omega &=& \gamma^{+0}_{0,0} \Psi^{\beta} \zeta_{\beta}
+ a_0 e_{\alpha\dot\alpha} Y^{\alpha} \zeta^{\dot\alpha}+h.c.,
 \\
\delta\Psi^{\alpha(2k)} &=&
{\tilde{\gamma}}^{0+}_{k,k}\Omega^{\alpha(2k)\dot\beta}
\zeta_{\dot\beta}
+ {\tilde{\gamma}}^{+0}_{k,k} \Omega^{\alpha(2k)\beta} \zeta_{\beta}
+ {\tilde{\gamma}}^{-0}_{k,k} \Omega^{\alpha(2k-1)} \zeta^{\alpha}
\nonumber \\
&& + {\tilde{\gamma}}^{0-}_{k,k}
\frac{\alpha^{-+}_{k+\iz}}{(2k+1)}e^{\alpha}{}_{\dot\alpha}
W^{\alpha(2k-1)} \zeta^{\dot\alpha}
+ {\tilde{\gamma}}^{0-}_{k,k} \alpha^{++}_{k-\iz}
e_{\beta\dot\alpha}W^{\alpha(2k)\beta}\zeta^{\dot\alpha}. \nonumber 
\end{eqnarray}
where the coefficient $a_0$ stands for:
\begin{eqnarray}
a_0=-(s+1)(\tilde{M}\pm\lambda)\sqrt{2(s-1)(\tilde{M}\mp\lambda)}C
\end{eqnarray}
At last, we have to consider remaining unfolded equations which
connect gauge sector with the sector of the gauge invariant 
zero-forms. The corresponding supertransformations have the form:
\begin{eqnarray}
\delta\Omega^{\alpha(2s-2)} &=&
\gamma^{0+}_{s-1,s-1}\Psi^{\alpha(2s-2)\dot\beta} \zeta_{\dot\beta}
+ \gamma^{+0}_{s-1,s-1} \Psi^{\alpha(2s-2)\beta} \zeta_{\beta}
+ \gamma^{-0}_{s-1,s-1} \Psi^{\alpha(2s-3)} \zeta^{\alpha}
\nonumber \\
&& + \gamma^{0-}_{s-1,s-1}
\frac{\tilde{\alpha}^{-+}_{s-\iz}}{(2s-1)}e^{\alpha}{}_{\dot\alpha}
Y^{\alpha(2s-3)} \zeta^{\dot\alpha}
+ \frac{\gamma^{0-}_{s-1,s-2}}{2} e_{\beta\dot\alpha}
Y^{\alpha(2s-2)\beta} \zeta^{\dot\alpha}
\nonumber \\
\delta W^{\alpha(s-1+m)\dot\alpha(s-1-m)} &=&
\gamma_{s-1,m}^{0-}Y^{\alpha(s+m-1)\dot\alpha(s-m-2)}
\zeta^{\dot\alpha} + \gamma_{s,m}^{-0} 
Y^{\alpha(s+m-2)\dot\alpha(s-1-m)}\zeta^{\alpha}
 \\
&& +
\frac{\gamma^{+0}_{s-1,m}}{\tilde{\alpha}^{++}_{s-\tz,m+\iz}}Y^{\alpha(s-1+m)\dot\alpha(s-1-m)\beta} \zeta_{\beta}
+ \frac{\gamma^{0+}_{s-1,m}}{\tilde{\alpha}^{++}_{s-\tz,m-\iz}}
Y^{\alpha(s-1+m)\dot\beta\dot\alpha(s-1-m)} \zeta_{\dot\beta}
\nonumber \\
\delta W^{\alpha(2s-2)} &=&
\frac{2\gamma^{+0}_{s-1,s-1}}{\tilde{\alpha}^{++}_{s-\tz}}
Y^{\alpha(2s-2)\beta} \zeta_{\beta}
+ \frac{\gamma^{0+}_{s-1,s-1}}{\tilde{\alpha}^{++}_{s-\tz,s-\tz}}
Y^{\alpha(2s-2)\dot\beta} \zeta_{\dot\beta}
 + \gamma_{s-1,s-1}^{-0} Y^{\alpha(2s-3)} \zeta^{\alpha} \nonumber
\end{eqnarray}
In particular, this gives us the relations between the constants
$C$, $\tilde{C}$ and $C_b$, $C_f$:
\begin{eqnarray}
C_b = \mp \frac{C}{\sqrt{2s(s-1)(\tilde{M}\pm s)}}, \qquad
C_f = \mp \tilde{C}\sqrt{2s(s-1)(\tilde{M}\pm s)}.
\end{eqnarray}
The parameters $C$, $\tilde{C}$ are restricted by the hermiticity
conditions only. Similarly to the massless case, their product
$C\tilde{C}$ is always imaginary. It is possible to restrict them
further by requiring the invariance of the sum of the bosonic and
fermionic Lagrangians. If one takes the normalization of the
Lagrangians as in \cite{KhZ19}, the connection between the parameters
is:
\begin{equation}
\tilde{C} = 4i\epsilon C
\end{equation}
One can see that this relation is in agreement with the hermiticity
conditions.

\subsubsection{Superblock $\tilde{s} = s + \iz$}

Now we repeat the same steps. The ansatz for the supertransformations
for the sector of gauge invariant zero-forms as well as the ansatz for
  the gauge sector are the same as before --- (\ref{ansatz_gi0f}) and
(\ref{ansatz_g0f}), (\ref{ansatz_g1f}) correspondingly. Hence, the
equations on the parameters of the supertransformations are also the
same (\ref{superblock_eqs1}), (\ref{superblock_eqs2}). But the
fermionic functions $\beta$, $\tilde{\beta}$ are different now and
this leads to the essentially different solution. For the sector of
the gauge invariant zero-forms we obtain for the bosonic functions 
$\delta$  $(m \ge 0)$:
\begin{eqnarray}
\delta^{+0}_{k,m} &=& C_b,
\nonumber \\
\delta^{0-}_{k,m} &=& \pm
\frac{(k-s)(\tilde{M}\pm(k+1)\lambda)}{(k-m)(k-m+1)}C_b,
\nonumber \\
\delta^{0+}_{k,m} &=& \mp \frac{(\tilde{M}\pm m\lambda)}{(s-m)}C_b,
\quad m > 0, \quad \delta^{0+}_{k,0} = \mp \epsilon C_b,
 \\
\delta^{-0}_{k,m} &=& \pm
\frac{(k-s)(\tilde{M}\pm(k+1)\lambda)}{(k+m)(k+m+1)}
\delta^{0+}_{k,m}, 
\nonumber 
\end{eqnarray}
and for the fermionic functions $\tilde{\delta}$ $(m \ge \iz)$:
\begin{eqnarray}
\tilde{\delta}^{+0}_{k+\iz,m+\iz} &=& (s-m)(s-m-1) C_f,
\nonumber \\
\tilde{\delta}^{0-}_{k+\iz,m+\iz} &=& \mp
\frac{(k+s+2)(\tilde{M}\mp(k+1)\lambda)}{(k-m)(k-m+1)}
\tilde{\delta}^{+0}_{k+\iz,m+\iz},
\nonumber \\
\tilde{\delta}^{0+}_{k+\iz,m+\iz} &=& \mp(s+m+1)
(\tilde{M}\mp m\lambda)C_f,
\\
\tilde{\delta}^{-0}_{k+\iz,m+\iz} &=& \mp
\frac{(k+s+2)(\tilde{M}\mp(k+1)\lambda)}{(k+m+1)(k+m+2)}
\tilde{\delta}^{0+}_{k+\iz,m+\iz}.
\nonumber 
\end{eqnarray}
Note that in this case $\delta^{0-}_{s,m} = 0$ as it should be. As in
the previous case, for $\lambda \ne 0$ we obtain a pair of additional
relations on the two arbitrary constants:
\begin{equation}
C^*_b = \pm \epsilon C_b, \qquad
C^*_f = \mp \epsilon C_f
\end{equation}

For the gauge sector supertransformation parameters $\gamma$ we
obtain $(m \ge 0)$:
\begin{eqnarray}
\gamma^{+0}_{k,m} &=& \pm \sqrt{k(s+k+2)(\tilde{M}\mp(k+1)\lambda)} C,
\quad k>0, 
\nonumber \\
\gamma^{+0}_{0,0} &=& \pm \sqrt{2(s+2)(\tilde{M}\mp\lambda)} C,
\nonumber \\
\gamma^{0-}_{k,m} &=& - 
\sqrt{\frac{(s-k)(\tilde{M}\pm(k+1)\lambda)}{k}}C,
\\
\gamma^{0+}_{k,m} &=& \mp
\frac{(s-m+1)(\tilde{M}\pm m\lambda)}{(k-m+1)(k-m+2)}
\gamma^{+0}_{k,m}, \quad m > 0, 
\nonumber \\
\gamma^{-0}_{k,m} &=& \mp
\frac{(s-m+1)(\tilde{M}\pm m \lambda)}{(k+m)(k+m+1)}\gamma^{0-}_{k,m},
\quad m> 0,
\nonumber
\end{eqnarray}
while for the parameters $\tilde{\gamma}$, correspondingly 
$(m \ge \iz)$:
\begin{eqnarray}
\tilde{\gamma}^{+0}_{k+\iz,m+\iz} &=& \pm
\sqrt{(k+1)(s-k-1)(\tilde{M}\pm(k+2)\lambda)}\tilde{C} ,
\nonumber \\
\tilde{\gamma}^{0-}_{k+\iz,m+\iz} &=& -
\sqrt{\frac{(s+k+2)(\tilde{M}\mp(k+1)\lambda)}{k}}\tilde{C},
\quad k>\iz,
\nonumber \\
\tilde{\gamma}^{0-}_{\iz,\iz} &=& -
\sqrt{\frac{(s+2)(\tilde{M}\mp\lambda)}{2}} \tilde{C},
\\
\tilde{\gamma}^{0+}_{k+\iz,m+\iz} &=& \mp
\frac{(s+m+1)(\tilde{M}\mp m \lambda)}{(k-m+1)(k-m+2)}
\tilde{\gamma}^{+0}_{k+\iz,m+\iz},
\nonumber \\
\tilde{\gamma}^{-0}_{k+\iz,m+\iz} &=& \mp
\frac{(s+m+1)(\tilde{M}\mp m\lambda)}{(k+m+1)(k+m+2)}
\tilde{\gamma}^{0-}_{k+\iz,m+\iz}.
\nonumber 
\end{eqnarray}
In the flat space $C$ and $\tilde{C}$ are the two arbitrary complex
constants while in $AdS$ $(\lambda \ne 0)$ they must satisfy the
relations similar to that of $C_b$ and $C_f$:
\begin{equation}
C^* = \pm \epsilon C, \qquad
\tilde{C}^* = \mp \epsilon \tilde{C}
\end{equation}

The supertransformations for the one-forms with $m=\pm k$ have to
contain zero-forms as well. The expressions for their
supertransformations are still given by (\ref{superblock_diagonal}),
but the expression for the coefficient $a_0$ is now:
\begin{equation}
a_0=-s(\tilde{M}\pm\lambda)\sqrt{2(s+2)(\tilde{M}\mp\lambda)}C
\end{equation}

At last let us turn to the remaining unfolded equations connecting two
sectors. In this case, it is fermionic fields supertransformations
which have to be modified: 
\begin{eqnarray}
\delta\Psi^{\alpha(2{\tilde{s}}-2)} &=& 
\tilde{\gamma}^{0+}_{{\tilde{s}}-1,{\tilde{s}}-1}\Omega^{\alpha(2{\tilde{s}}-2)\dot\beta} \zeta_{\dot\beta}
+
\tilde{\gamma}^{+0}_{{\tilde{s}}-1,{\tilde{s}}-1}\Omega^{\alpha(2{\tilde{s}}-2)\beta} \zeta_{\beta}
+
\tilde{\gamma}^{-0}_{{\tilde{s}}-1,{\tilde{s}}-1}\Psi^{\alpha(2{\tilde{s}}-3)} \zeta^{\alpha}
\nonumber \\
&& + \tilde{\gamma}^{0-}_{{\tilde{s}}-1,{\tilde{s}}-1}
\frac{\alpha^{-+}_{{\tilde{s}}-\iz}}{(2{\tilde{s}}-1)}
e^{\alpha}{}_{\dot\alpha} W^{\alpha(2s-3)} \zeta^{\dot\alpha}
+ \frac{\tilde{\gamma}^{0-}_{{\tilde{s}}-1,{\tilde{s}}-2}}{2}
e_{\alpha\dot\alpha} W^{\alpha(2{\tilde{s}}-1)} \zeta^{\dot\alpha}
\nonumber \\
\delta Y^{\alpha({\tilde{s}}-1+m)\dot\alpha({\tilde{s}}-1-m)} &=&
\tilde{\gamma}_{{\tilde{s}}-1,m}^{-0}
W^{\alpha({\tilde{s}}+m-2)\dot\alpha({\tilde{s}}-1-m)} \zeta^{\alpha}
+ \tilde{\gamma}_{{\tilde{s}}-1,m}^{0-}
W^{\alpha({\tilde{s}}+m-1)\dot\alpha({\tilde{s}}-m-2)}
\zeta^{\dot\alpha}
 \\
&& +
\frac{\tilde{\gamma}^{0+}_{{\tilde{s}}-1,m}}{\alpha^{++}_{{\tilde{s}}-\tz,m-\iz}} 
W^{\alpha({\tilde{s}}-1+m)\beta\dot\alpha({\tilde{s}}-1-m)}\zeta_{\beta}
+
\frac{\tilde{\gamma}^{+0}_{{\tilde{s}}-1,m}}{\alpha^{++}_{{\tilde{s}}-\tz,m+\iz}}
W^{\alpha({\tilde{s}}-1+m)\dot\alpha({\tilde{s}}-1-m)\dot\beta}\zeta_{\dot\beta}
\nonumber \\
\delta Y^{\alpha(2{\tilde{s}}-2)} &=& 
\frac{2\tilde{\gamma}^{+0}_{{\tilde{s}}-1,{\tilde{s}}-1}}{\alpha^{++}_{{\tilde{s}}-\tz}} W^{\alpha(2{\tilde{s}}-2)\beta} \zeta_{\beta}
+
\frac{\tilde{\gamma}^{0+}_{{\tilde{s}}-1,{\tilde{s}}-1}}{\alpha^{++}_{{\tilde{s}}-\tz,{\tilde{s}}-\tz}} 
W^{\alpha(2{\tilde{s}}-2)\dot\beta} \zeta_{\dot\beta}
+
\tilde{\gamma}_{{\tilde{s}}-1,{\tilde{s}}-1}^{-0}W^{\alpha(2{\tilde{s}}-3)} \zeta^{\alpha} \nonumber
\end{eqnarray}
For the consistency the constants $C$, $\tilde{C}$ have to be
connected with the constants $C_b$, $C_f$ as follows:
\begin{equation}
C_b = \pm C\sqrt{(s-1)(2s+1)(\tilde{M}\mp s)}, \qquad
C_f = \pm\frac{\tilde{C}}{\sqrt{(s-1)(2s+1)(\tilde{M}\mp s)}}
\end{equation}
Apart from the hermiticity conditions, the constants $C$ and
$\tilde{C}$ are arbitrary. If the sum of the Lagrangians is required
to be invariant, these constants turn out to be connected:
\begin{equation}
\tilde{C} = 4i\epsilon C
\end{equation}
Again, this relation is in agreement with the hermiticity conditions.

\subsection{Supermultiplets}

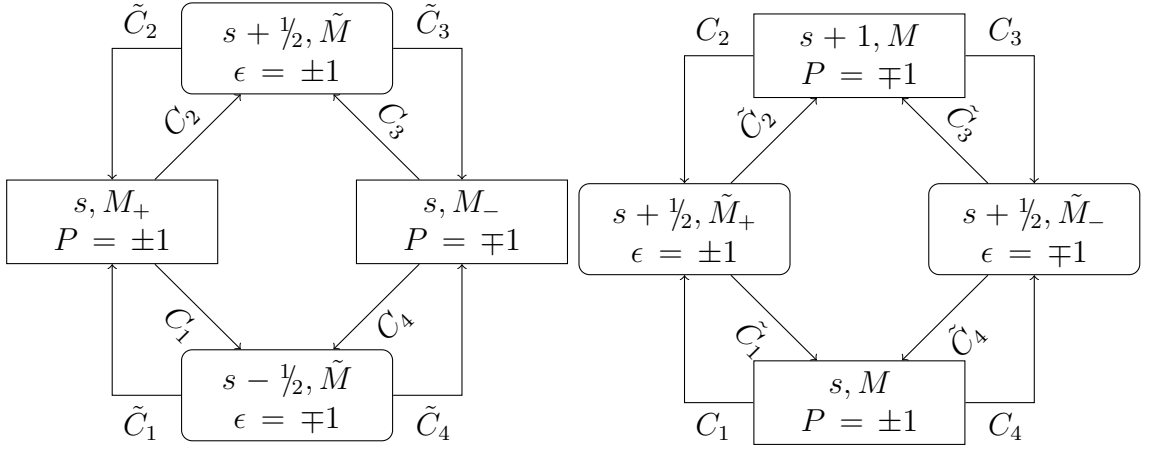
\begin{figure}[h]
\centering
\begin{tikzpicture}
\draw
(0,-2.3) node(s)[fblock,text width=2.5cm,align=center]
{$s-\iz,\tilde{M}$\\$\epsilon=\mp1$}
(-2.3,0) node(s1)[bblock,text width=2.5cm,align=center]
{$s,M_+$\\$P=\pm1$}
(2.3,0) node(s2)[bblock,text width=2.5cm,align=center]
{$s,M_-$\\$P=\mp1$}
(0,2.3) node(s3)[fblock,text width=2.5cm,align=center]
{$s+\iz,\tilde{M}$\\$\epsilon=\pm1$};

\draw[->](s.west) -| (s1.south) node [below right,midway]
{$\tilde{C}_1$};
\draw[->](s1.315) -- (s.135) node [midway,below,sloped] {$C_1$};
\draw[->](s.east) -| (s2.south) node [below left,midway]
{$\tilde{C}_4$};
\draw[->](s2.225) -- (s.45) node [midway,below,sloped] {$C_4$};

\draw[->](s3.west) -| (s1.north) node [above right,midway]
{$\tilde{C}_2$};
\draw[->](s1.45) -- (s3.225) node [midway,above,sloped] {$C_2$};
\draw[->](s3.east) -| (s2.north) node [above left,midway]
{$\tilde{C}_3$};
\draw[->](s2.135) -- (s3.315) node [midway,above,sloped] {$C_3$};
\end{tikzpicture}
\begin{tikzpicture}
\draw
(0,-2.3) node(s)[bblock,text width=2.5cm,align=center]
{$s,M$\\$P=\pm1$}
(-2.3,0) node(s1)[fblock,text width=2.5cm,align=center]
{$s+\iz,\tilde{M}_+$\\ $\epsilon=\pm1$}
(2.3,0) node(s2)[fblock,text width=2.5cm,align=center]
{$s+\iz,\tilde{M}_-$\\ $\epsilon=\mp1$}
(0,2.3) node(s3)[bblock,text width=2.5cm,align=center]
{$s+1,M$\\$P=\mp1$};

\draw[->](s.west) -| (s1.south) node [below right,midway] {$C_1$};
\draw[->](s1.315) -- (s.135) node [midway,below,sloped]
{$\tilde{C}_1$};
\draw[->](s.east) -| (s2.south) node [below left,midway] {$C_4$};
\draw[->](s2.225) -- (s.45) node [midway,below,sloped]
{$\tilde{C}_4$};

\draw[->](s3.west) -| (s1.north) node [above right,midway] {$C_2$};
\draw[->](s1.45) -- (s3.225) node [midway,above,sloped]
{$\tilde{C}_2$};
\draw[->](s3.east) -| (s2.north) node [above left,midway] {$C_3$};
\draw[->](s2.135) -- (s3.315) node [midway,above,sloped]
{$\tilde{C}_3$};
\end{tikzpicture}
\caption{Structure of massive HS supermultiplets. Sharp boxes
represent bosons, while the skew ones represent fermions. The letter
$P$ is boson parity. Each arrow from A to B corresponds to the terms
with B fields in the variation of A fields under the
supertransformation. All such terms are proportional to their own
constant $C_i$ (resp. $\tilde{C}_i$). The parameters $M_+,M_-$ (resp.
$\tilde{M}_+,\tilde{M}_-$) are the roots of
$M^2=\tilde{M}(\tilde{M}\pm \lambda)$ with a sign chosen respectively;
note that $\tilde{M}_--\tilde{M}_+=\lambda$. }
\label{fig:hsm_structure}
\end{figure}

We build the supermultiplets now. A massive supermultiplet contains
two bosons and two fermions; each pair of one boson and one fermion
forms a superblock. It was shown in \cite{BKhSZ19} that the bosons
have the opposite parity and the fermions have opposite mass terms
sign. This leaves four possible structures of the supermultiplet, as
shown in the Figure \ref{fig:hsm_structure}.
Each pair of fields connected by a pair of arrows forms a superblock. 
One can see that the commutator of two supertransformations
transforms a field into a combination of two fields and one of these
fields corresponds to another particle. The coefficients $C_i$ and
$\tilde{C}_i$ have to be tuned to get rid of such terms. This gives
certain equalities for the products $C_i\tilde{C}_i$. The rest of the
terms must form the transformations of the $AdS$ algebra. Again, we
consider integer and half-integer superspin (i.e. average spin of the
supermultiplet $\langle s \rangle$) cases separately.  

\subsubsection{Integer superspin case}

In case of integer superspin, the coefficients $C_i$ and $\tilde{C}_i$
mus satisfy:
\begin{equation}
\label{coeff_products}
C_1\tilde{C}_1 = - C_2\tilde{C}_2 = C_3\tilde{C}_3 = - C_4\tilde{C}_4
= iC^2, \qquad C_1C_3 = C_2C_4, \qquad \tilde{C}_1\tilde{C}_3
=\tilde{C}_2\tilde{C}_4
\end{equation}
If one also requires the invariance of the sum of the Lagrangians for
all four members, the coefficients become fixed up to a single scale
factor $C$. If the highest-spin fermion has $\epsilon=1$, the
constants are:
\begin{eqnarray}
C_1 &=& \frac{C}{2}, \qquad C_2=\frac{C}{2}, \qquad
C_3 = i\frac{C}{2}, \qquad C_4 = i\frac{C}{2},
\nonumber \\
\tilde{C}_1 &=& 2iC, \qquad \tilde{C}_2 = -2iC, \qquad
\tilde{C}_3 = 2C, \qquad \tilde{C}_4 = -2C.
\end{eqnarray}
If the highest-spin fermion has $\epsilon=-1$, the constants are:
\begin{eqnarray}
C_1 &=& -i\frac{C}{2}, \qquad C_2 = -i\frac{C}{2}, \qquad
C_3 = \frac{C}{2}, \qquad C_4 = \frac{C}{2},
\nonumber \\
\tilde{C}_1 &=& -2C, \qquad \tilde{C}_2 = 2C, \qquad
\tilde{C}_3 = 2iC, \qquad \tilde{C}_4 = -2iC.
\end{eqnarray}
We give the resulting expression for the commutator for the bosonic
field $\Omega^{\alpha(k+m)\dot\alpha(k-m)}$ as an example:
\begin{eqnarray}
[\delta_1,\delta_2]\Omega^{\alpha(k+m)\dot\alpha(k-m)} &=&
4 i C^2 \tilde{M} (\langle s \rangle+\iz)
\nonumber \\
&=& \bigg[
\lambda \Omega^{\alpha(k+m)\dot\alpha(k-m-1)\dot\beta}
\eta_{\dot\beta}{}^{\dot\alpha} + \lambda
\Omega^{\alpha(k+m-1)\beta\dot\alpha(k-m)} \eta_{\beta}{}^{\alpha}
\nonumber \\
&& + \alpha^{-+}_{k,m}
\Omega^{\alpha(k+m-1)\dot\alpha(k-m)\dot\beta}
\xi^\alpha{}_{\dot\beta} +
\Omega^{\alpha(k+m)\beta\dot\alpha(k-m-1)}\xi_\beta{}^{\dot\alpha}
\nonumber \\
&& + \alpha^{--}_{k,m}
\Omega^{\alpha(k+m-1)\dot\alpha(k-m-1)}\xi^{\alpha\dot\alpha} +
\alpha^{++}_{k,m}
\Omega^{\alpha(k+m)\beta\dot\alpha(k-m)\dot\beta}
\xi_{\beta\dot\beta} \bigg]
\end{eqnarray}
Recall that
\begin{equation}
\eta^{\alpha(2)} = 2{\zeta_1}^\alpha{\zeta_2}^\alpha, \qquad
\eta^{\dot\alpha(2)} =
2{\zeta_1}^{\dot\alpha}{\zeta_2}^{\dot\alpha},\qquad
\xi^{\alpha\dot\alpha} = {\zeta_1}^\alpha
{\zeta_2}^{\dot\alpha} - {\zeta_1}^{\dot\alpha}{\zeta_2}^\alpha
\end{equation}
The factor $4iC^2\tilde{M}(\langle s \rangle+\iz)$ is the same for all
fields. The coefficients $\alpha^{ij}_{k,m}$ correspond to the
same particle as the field $\Omega^{\alpha(k+m)\dot\alpha(k-m)}$. By
comparing the expression with then unfolded equations, one can see
that it is indeed a combination of pseudotranslations and Lorentz
transformations. 

\subsubsection{Half-integer integer superspin case}

In case of half-integer superspin, the products of the coefficients
$C_i$ and $\tilde{C}_i$ are fixed by the same relations
(\ref{coeff_products}).  The requirement of the invariance for the sum
of the Lagrangians fixes the coefficients up to the single scale
factor. In case of even-parity highest-spin boson, the coefficients
are:
\begin{eqnarray}
C_1 &=& i\frac{C}{2}, \qquad C_2 = \frac{C}{2}, \qquad
C_3 = \frac{C}{2}, \qquad C_4 = i\frac{C}{2},
\nonumber \\
\tilde{C}_1 &=& 2C, \qquad \tilde{C}_2 = -2iC, \qquad
\tilde{C}_3 = 2iC, \qquad \tilde{C}_4 = -2C.
\end{eqnarray}
If the highest-spin boson is parity-odd, the constants are:
\begin{eqnarray}
C_1 &=& \frac{C}{2}, \qquad C_2 = -i\frac{C}{2}, \qquad
C_3 = -i\frac{C}{2}, \qquad C_4 = \frac{C}{2},
\nonumber \\
\tilde{C}_1 &=& 2iC, \qquad \tilde{C}_2 = 2C, \qquad
\tilde{C}_3 = -2C, \qquad \tilde{C}_4 = -2iC.
\end{eqnarray}
Again, we present a commutator of the supertransformations for the
field $\Omega^{\alpha(k+m)\dot\alpha(k-m)}$ as an example:
\begin{eqnarray}
[\delta_1,\delta_2]\Omega^{\alpha(k+m)\dot\alpha(k-m)} &=&
2 i C^2 (\tilde{M}_++\tilde{M}_-) (\langle s \rangle+\iz)
\nonumber \\
&=& \bigg[ \lambda
\Omega^{\alpha(k+m)\dot\alpha(k-m-1)\dot\beta}\eta_{\dot\beta}{}^{\dot\alpha} + \lambda
\Omega^{\alpha(k+m-1)\beta\dot\alpha(k-m)} \eta_{\beta}{}^{\alpha}
\nonumber \\\
&& + \alpha^{-+}_{k,m}
\Omega^{\alpha(k+m-1)\dot\alpha(k-m)\dot\beta}\xi^\alpha{}_{\dot\beta}
+ \Omega^{\alpha(k+m)\beta\dot\alpha(k-m-1)}\xi_\beta{}^{\dot\alpha}
\nonumber \\
&& + \alpha^{--}_{k,m}
\Omega^{\alpha(k+m-1)\dot\alpha(k-m-1)}\xi^{\alpha\dot\alpha} +
\alpha^{++}_{k,m}
\Omega^{\alpha(k+m)\beta\dot\alpha(k-m)\dot\beta}
\xi_{\beta\dot\beta} \bigg]
\end{eqnarray}
One can see that the structure of the commutator is the same as in the
previous case. The factor 
$2 i C^2 (\tilde{M}_++\tilde{M}_-) (\langle s \rangle+\iz)$ is
slightly different now. Again, it is the same for all fields. The
coefficients $\alpha^{ij}_{k,m}$ correspond to the same particle as
the field $\Omega^{\alpha(k+m)\dot\alpha(k-m)}$.

\section{Infinite spin supermultiplets}

Recently it became clear that the gauge invariant formalism we use for
the description of massive higher spin fields nicely works for the
infinite spin limit as well
\cite{Met16,Met17,Zin17,KhZ17,Met18,KhZ19}. Moreover, the first
examples of the infinite spin supermultiplets in the flat space were
constructed \cite{Zin17,BKhSZ19b} (see also recent paper
\cite{Naj19}). In this section we consider unfolded formulation of the
infinite spin supermultiplets both in the flat and $AdS_4$ spaces.
These two cases turns out to be rather different, so we consider them
separately in the two subsequent subsections.

Let us begin with the general considerations. In the infinite spin
limit the gauge invariant formulation does not contain any gauge
invariant zero-forms so we have the gauge one-forms 
$\Omega$, $\Psi$ and Stueckelberg zero-forms $W$, $Y$ only. In this,
the unfolded equations is just the infinite set of the zero-curvature
conditions:
\begin{eqnarray}
\mathcal{R}^{\alpha(k+m)\dot\alpha(k-m)} &=& 0, \qquad
\mathcal{C}^{\alpha(k+m)\dot\alpha(k-m)} = 0
\nonumber \\
\mathcal{F}^{\alpha(k+m)\dot\alpha(k-m)} &=& 0, \qquad
\mathcal{D}^{\alpha(k+m)\dot\alpha(k-m)} = 0
\end{eqnarray}
The expressions for the bosonic curvatures $\mathcal{R}$ and 
$\mathcal{C}$ are still given by (\ref{curvatures1}),
(\ref{curvatures2}), while the fermionic ones are still defined by
(\ref{curvatures1a}), (\ref{curvatures2a}) but with different
functions $\alpha$, $\tilde{\alpha}$ (see below). Similarly, the
general ansatz for the supertransformations for the Stueckelberg
zero-forms is still (\ref{ansatz_g0f}) and for the one-forms is still
(\ref{ansatz_g1f}) and (\ref{superblock_diagonal}). 

\subsection{Flat space}

In the infinite spin limit the gauge invariant formalism leads to the
massless and tachyonic solutions for bosons and only massless ones for
fermions (because the tachyonic ones are non unitary) 
\cite{Met16,Met17,KhZ19}. This leaves us the only possibility --- a
massless infinite spin supermultiplet in agreement with the
classification in \cite{BKRX02}. 

\begin{figure}[ht!]
\centering
\begin{tikzpicture}
\draw
(0,-2.3) node(s)[fblock,text width=2.5cm,align=center]
{$\epsilon=\mp1$}
(-2.3,0) node(s1)[bblock,text width=2.5cm,align=center] {$P=\pm1$}
(2.3,0) node(s2)[bblock,text width=2.5cm,align=center] {$P=\mp1$}
(0,2.3) node(s3)[fblock,text width=2.5cm,align=center]
{$\epsilon=\pm1$};

\draw[->](s.west) -| (s1.south) node [below right,midway]
{$\tilde{C}_1$}
node [above right,midway] {$-$};
\draw[->](s1.315) -- (s.135) node [midway,below,sloped] {$C_1$};
\draw[->](s.east) -| (s2.south) node [below left,midway]
{$\tilde{C}_4$}
node [above left,midway] {$+$};
\draw[->](s2.225) -- (s.45) node [midway,below,sloped] {$C_4$};

\draw[->](s3.west) -| (s1.north) node [above right,midway]
{$\tilde{C}_2$}
node [below right,midway] {$+$};
\draw[->](s1.45) -- (s3.225) node [midway,above,sloped] {$C_2$};
\draw[->](s3.east) -| (s2.north) node [above left,midway]
{$\tilde{C}_3$}
node [below left,midway] {$-$};
\draw[->](s2.135) -- (s3.315) node [midway,above,sloped] {$C_3$};
\end{tikzpicture}
\caption{Structure of the infinite spin supermultiplet. The parameters
$\mu$ are equal for each particle and are omitted. Again, bosons are
represented by sharp boxes, while the fermions - by rounded ones. The
sign choice for the each superblock is now indicated in the
corresponding corner of the picture}
\label{fig:fs_issm}
\end{figure}
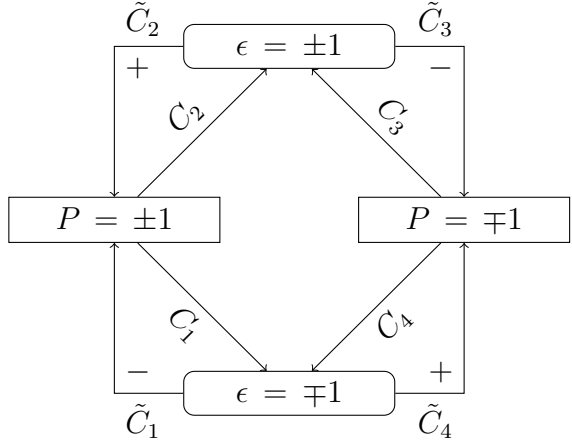

For the massless infinite spin boson the functions $\alpha$ have a
rather simple form:
\begin{eqnarray}
\alpha^{++}_{k,m} &=& \frac{\sqrt{k(k+1)}\mu}{(k-m+1)(k-m+2)}
\nonumber \\
\alpha^{-+}_{k,m} &=& \frac{\mu^2}{(k+m)(k+m+1)(k-m+1)(k-m+2)}
\nonumber \\
\alpha^{-+}_{k,0} &=& 1 
\\
\alpha^{--}_{k,m} &=& \frac{\mu}{(k+m)(k+m+1)\sqrt{k(k-1)}} 
\nonumber
\end{eqnarray}
where $\mu$ is a dimensionful parameter related with the eigenvalue
of the second Casimir operator of Poincare group. Similarly, for the
massless infinite spin fermions we have:
\begin{eqnarray}
\tilde{\alpha}^{++}_{k,m} &=& \frac{(k+1)\tilde{\mu}}{(k-m+1)(k-m+2)}
\nonumber \\
\tilde{\alpha}^{-+}_{k,m} &=& 
\frac{\tilde{\mu}^2}{(k-m+1)(k-m+2)(k+m+1)(k+m+2)}
\nonumber \\
\tilde{\alpha}^{-+}_{k,0} &=& \epsilon \frac{\tilde{\mu}}{(k+1)(k+2)},
\qquad \epsilon = \pm 1 \\
\tilde{\alpha}^{--}_{k,m} &=& \frac{\tilde{\mu}}{(k+m+1)(k+m+2)k}
\nonumber
\end{eqnarray}
{\bf Superblock} Let us consider a superblock containing one such
boson and one fermion. First of all, supersymmetry requires that their
dimensionfull parameters must be equal $\mu = \tilde{\mu}$. Then we
obtain the following expressions for the parameters of the
supertransformations for the boson:
\begin{eqnarray}
\gamma^{+0}_{k,m} &=& \sqrt{k} C 
\nonumber \\
\gamma^{-0}_{k,m} &=& \frac{\mu}{(k+m)(k+m+1)\sqrt{k}} C
\nonumber \\
\gamma^{0+}_{k,m} &=& - \epsilon \frac{\sqrt{k}\mu}{(k-m+1)(k-m+2)}
C^*
\\
\gamma^{0-}_{k,m} &=& - \epsilon \frac{1}{\sqrt{k}} C^* \nonumber
\end{eqnarray}
and for the fermion:
\begin{eqnarray}
\tilde{\gamma}^{+0}_{k,m} &=& \sqrt{(k+1)} \tilde{C} 
\nonumber \\
\tilde{\gamma}^{-0}_{k,m} &=& \frac{\mu}{(k+m+1)(k+m+2)\sqrt{k}}
\tilde{C}
\nonumber \\
\tilde{\gamma}^{0+}_{k,m} &=& \epsilon 
\frac{\sqrt{(k+1)}\mu}{(k-m+1)(k-m+2)} \tilde{C}^*
 \\
\tilde{\gamma}^{0-}_{k,m} &=& \epsilon \frac{1}{\sqrt{k}} \tilde{C}^*
\nonumber
\end{eqnarray}
Here $C$ and $\tilde{C}$ are two arbitrary complex constants. It is
easy to check that the algebra of these supertransformations is not
closed so to construct a supermultiplet we have to consider a pair of
bosons and a pair of fermions. \\
{\bf Supermultiplet} In the flat space, there exists only one infinite
spin supermultiplet, with its structure shown in the Figure
\ref{fig:fs_issm}.
As in the Lagrangian formulation \cite{BKhSZ19b},
we have found that the two bosons must have opposite parity, while the
two fermions must have opposite signs of the mass-like terms
$\epsilon_2 = - \epsilon_1$. Moreover, all the products 
$C_i\tilde{C}_i$, $i=1,2,3,4$ must be imaginary and satisfy the
following relations:
\begin{eqnarray}
C_1\tilde{C}_1 &=& - C_2\tilde{C}_2 = C_3\tilde{C}_3 
= - C_4\tilde{C}_4
\nonumber \\
C_2\tilde{C}_3 &=& - C_1\tilde{C}_4, \qquad
C_3\tilde{C}_4 = - C_2\tilde{C}_1.
\end{eqnarray}
For definiteness, we assume that the first boson is parity-even, and
the first fermion has $\epsilon_1 = 1$. If we also require that not
only unfolded equations but also the sum of the four Lagrangians is
invariant under the supertransformations we obtain
\begin{eqnarray}
C_1 &=& \frac{C}{2}, \qquad C_2 = \frac{C}{2}, \qquad 
C_3 = i\frac{C}{2}, \qquad C_4 = i\frac{C}{2},
\nonumber \\
\tilde{C}_1 &=& 2iC, \qquad  \tilde{C}_2 = -2iC, \qquad 
\tilde{C}_3 = 2C, \qquad \tilde{C}_4 = -2C.
\end{eqnarray}
Once again, we provided as an example the explicit expressions for
the commutator of the two supertransformations on the one-form 
$\Omega$:
\begin{eqnarray}
[\delta_1,\delta_2]\Omega^{\alpha(k+m)\dot\alpha(k-m)} &=& 2 i C^2
\bigg[ \alpha^{-+}_{k,m} 
\Omega^{\alpha(k+m-1)\dot\alpha(k-m)\dot\beta} 
\xi^\alpha{}_{\dot\beta} +
\Omega^{\alpha(k+m)\beta\dot\alpha(k-m-1)}\xi_\beta{}^{\dot\alpha}
\nonumber \\
&& + \alpha^{--}_{k,m}
\Omega^{\alpha(k+m-1)\dot\alpha(k-m-1)}\xi^{\alpha\dot\alpha} +
\alpha^{++}_{k,m}
\Omega^{\alpha(k+m)\beta\dot\alpha(k-m)\dot\beta}
\xi_{\beta\dot\beta} \bigg]
\end{eqnarray}

\subsection{$AdS_4$ space}

In this case for the infinite spin limit the gauge invariant formalism
provides a whole range of the unitary solutions both for the bosons as
well as for the fermions \cite{Met16,Met17,KhZ19}. But as we have
already noted for the construction of the supermultiplets it is
crucial to have a factorization of the main functions $\alpha^{-+}$
and $\tilde{\alpha}^{-+}$. The only such possibility we have found ---
so called "partially massless" infinite spin particles when the
spectrum of helicities is $s \le |h| < \infty$, where integer or 
half-integer $s$ denotes the lowest helicity. In this case the main
functions look very similar to the massive finite spin case:
\begin{eqnarray}
\alpha^{-+}_m &=& (m-s-1)(m+s) [m(m-1)\lambda^2 - M^2]
\nonumber \\
\tilde{\alpha}^{-+}_m &=& (m-\tilde{s}-1)(m+\tilde{s})
[(m-\iz)^2\lambda^2 - \tilde{M}^2]
\end{eqnarray}
Moreover, it appears that the bosonic and fermionic mass parameters
must still satisfy the same relation 
$M^2 = \tilde{M}[\tilde{M} \pm\lambda]$. As a result, we obtain:
\begin{equation}
\alpha^{-+}_m = (m-s-1)(m+s)[m\lambda \pm \tilde{M}]
[(m-1)\lambda \mp \tilde{M}]
\end{equation}
As in the massive case, we begin with the construction of two possible
superblocks with $\tilde{s} = s \pm \iz$. \\
{\bf Superblock $\tilde{s} = s - \iz$} For the bosonic functions
$\gamma$ we obtain ($k \ge s$, $m \ge 0$):
\begin{eqnarray}
\gamma^{+0}_{k,m} &=& \sqrt{k(k+1-s)((k+1)\lambda\mp\tilde{M})} C,
\nonumber \\
\gamma^{0-}_{k,m} &=& 
\sqrt{\frac{(k+s+1)((k+1)\lambda\pm\tilde{M})}{k}}C,
\\
\gamma^{0+}_{k,m} &=& 
\frac{(s+m)(\tilde{M}\pm m\lambda)}{(k-m+1)(k-m+2)}\gamma^{+0}_{k,m},
\qquad m > 0, \nonumber \\
\gamma^{-0}_{k,m} &=& 
\frac{(s+m)(\tilde{M}\pm m \lambda)}{(k+m)(k+m+1)} \gamma^{0-}_{k,m},
\qquad m> 0, \nonumber 
\end{eqnarray}
while for the fermionic functions $\tilde{\gamma}$ 
($k \ge \tilde{s}$, $m \ge \iz$):
\begin{eqnarray}
\tilde{\gamma}^{+0}_{k+\iz,m+\iz} &=& 
\sqrt{(k+1)(s+k+2)((k+2)\lambda\pm\tilde{M})} \tilde{C} 
\nonumber \\
\tilde{\gamma}^{0-}_{k+\iz,m+\iz} &=& 
\sqrt{\frac{(k+1-s)((k+1)\lambda\mp\tilde{M})}{k}} \tilde{C}
\nonumber \\
\tilde{\gamma}^{0+}_{k+\iz,m+\iz} &=& 
\frac{(s-m)(\tilde{M}\mp m \lambda)}{(k-m+1)(k-m+2)}
\tilde{\gamma}^{+0}_{k+\iz,m+\iz},
 \\
\tilde{\gamma}^{-0}_{k+\iz,m+\iz} &=& 
\frac{(s-m)(\tilde{M}\mp m\lambda)}{(k+m+1)(k+m+2)}
\tilde{\gamma}^{0-}_{k+\iz,m+\iz}, \nonumber
\end{eqnarray}
Since $\lambda \ne 0$, we obtain also a pair of relations on these two
parameters $C$ and $\tilde{C}$:
\begin{equation}
C^* = \mp \epsilon C, \qquad
\tilde{C}^* = \pm \epsilon \tilde{C}
\end{equation}
At the same time, the relation between $c$ and $\tilde{C}$ from the
invariance for the  sum of the two Lagrangians appears to be different
form the massive case:
\begin{equation}
\tilde{C} = \pm 4i\epsilon C
\end{equation}
and this turns out to be important (see below).

\noindent
{\bf Superblock $\tilde{s} = s + \iz$} In this case the bosonic
functions $\gamma^{ij}_{k,m}$ are ($k \ge s$, $m \ge 0$):
\begin{eqnarray}
\gamma^{+0}_{k,m} &=& \sqrt{k(s+k+2)((k+1)\lambda\mp\tilde{M})} C,
\nonumber \\
\gamma^{0-}_{k,m} &=&
\sqrt{\frac{(k-s)((k+1)\lambda\pm\tilde{M})}{k}}C
\\
\gamma^{0+}_{k,m} &=& 
\frac{(s-m+1)(\tilde{M}\pm m\lambda)}{(k-m+1)(k-m+2)} 
\gamma^{+0}_{k,m}, \qquad m > 0, 
\nonumber \\
\gamma^{-0}_{k,m} &=& 
\frac{(s-m+1)(\tilde{M}\pm m \lambda)}{(k+m)(k+m+1)} 
\gamma^{0-}_{k,m}, \qquad m> 0,
\nonumber 
\end{eqnarray}
and for the fermionic ones $\tilde{\gamma}$ ($k \ge \tilde{s}$, 
$m \ge \iz$):
\begin{eqnarray}
\tilde{\gamma}^{+0}_{k+\iz,m+\iz} &=& 
\sqrt{(k+1)(k+1-s)((k+2)\lambda\pm\tilde{M})} \tilde{C}, 
\nonumber \\
\tilde{\gamma}^{0-}_{k+\iz,m+\iz} &=& \pm
\sqrt{\frac{(s+k+2)((k+1)\lambda\mp\tilde{M})}{k}} \tilde{C},
\nonumber \\
\tilde{\gamma}^{0+}_{k+\iz,m+\iz} &=& \mp
\frac{(s+m+1)(\tilde{M}\mp m \lambda)}{(k-m+1)(k-m+2)}
\tilde{\gamma}^{+0}_{k+\iz,m+\iz}
 \\
\tilde{\gamma}^{-0}_{k+\iz,m+\iz} &=& \mp
\frac{(s+m+1)(\tilde{M}\mp m\lambda)}{(k+m+1)(k+m+2)}
\tilde{\gamma}^{0-}_{k+\iz,m+\iz},
\nonumber 
\end{eqnarray}
In this case we also obtain
\begin{equation}
C^* = \pm \epsilon C, \qquad
\tilde{C}^* = \mp \epsilon \tilde{C}
\end{equation}
Again, the relation between $C$ and $\tilde{C}$ which follows from the
Lagrangians invariance, is slightly different:
\begin{equation}
\tilde{C} = \mp 4i\epsilon C
\end{equation}

\noindent
{\bf Supermultiplets} 
Similarly to the massive case, there exist two different solutions for
the infinite spin supermultiplet in $AdS_4$, which resemble those with
the integer superspin and half integer superspin. Their structure is
the same as in the massive case (see Figure \ref{fig:hsm_structure}).
The coefficients $C_i,\tilde{C}_i$ are restricted by the same
conditions as in (\ref{coeff_products}):
\begin{equation}
C_1\tilde{C}_1 = - C_2\tilde{C}_2 = C_3\tilde{C}_3 = C_4\tilde{C}_4 =
iC^2, \qquad C_1C_3 = C_2C_4, \qquad
\tilde{C}_1\tilde{C}_3 = \tilde{C}_2\tilde{C}_4
\end{equation}
The expressions for the commutators are also the same as in the
massive supermultiplet case. However, the restrictions following from
the Lagrangian invariance cannot be satisfied, as they require, for
instance, the bosons to have the same parity. A possible way to
restore the invariance is to change the sign of one bosonic and one
fermionic Lagrangians so that the connection between $C_i$ and 
$\tilde{C}_i$ becomes $\tilde{C}_i=4i\epsilon C$ as in in the massive
case. But this spoils the unitarity of the theory and this resembles
the situation with the non-unitary partially massless finite spin
supermultiplets constructed in \cite{BKhSZ19a}.

\section*{Conclusion}

In this paper we have constructed the unfolded formulation for the
massive higher spin $N=1$ supermultiplets in $AdS_4$. Our results are
in complete agreement with the results of \cite{BKhSZ19} where the
Lagrangian formulation of such supermultiplets were developed. We also
consider an infinite spin limit for these supermultiplets with the
results also consistent with that of \cite{BKhSZ19a}.

\section*{Acknowledgements}
Authors are grateful to the I. L. Buchbinder and T. V. Snegirev for
collaboration. M.Kh. is grateful to Foundation for the Advancement of
Theoretical Physics and Mathematics "BASIS" for their support of the
work.

\appendix

\section{Notations and conventions}

In the paper, we adopt the "condensed notation" of the indices.
Namely, if an expression contains $n$ consecutive indices, denoted by
the same letter with different indices (e.g. $\alpha_1,\alpha_2,\ldots
\alpha_n$) and is symmetric on them, we simply write the letter, with
the number $n$ in parentheses if $n>1$ (e.g. $\alpha(n)$). For
example:
\begin{equation}
\Phi^{\alpha_1,\alpha_2,\alpha_3} = \Phi^{\alpha(3)}, \qquad
\zeta^{\alpha_1} \Omega^{\alpha_2\alpha_3} = \zeta^\alpha
\Omega^{\alpha(2)}
\end{equation}
We define symmetrization over indices as the sum of the minimal number
of terms necessary without normalization multiplier.

We use the multispinor formalism in four dimensions as in the paper
\cite{DS14}. Every vector index is transformed into a pair of spinor
indices: $V^\mu\sim V^{\alpha,\dot{\alpha}}$, where
$\alpha,\dot{\alpha}=1,2$. Dotted and undotted indices are transformed
into one another under the hermitian conjugation:
\begin{equation}
\left(\Omega^{\alpha{\dot{\alpha}(2)}}\right)^\dagger=\Omega^{\alpha(2){\dot{\alpha}}}
\end{equation}
The spin-tensors, i.e. fields with odd number of indices, are
Grassmannian. For example,
\begin{equation}
A^{\alpha(2)\dot\alpha} \eta^{\alpha} = - \eta^{\alpha}
A^{\alpha(2)\dot\alpha}
\end{equation}
Under the hermitian conjugation, the order of fields is reversed:
\begin{equation}
\left(A^{\alpha(2)\dot\alpha}\eta^{\alpha}\right)^\dagger
= \eta^{\alpha} A^{\alpha(2)\dot\alpha} = -
A^{\alpha(2)\dot\alpha}\eta^{\alpha}
\end{equation}
The metrics for the spinor indices is an antisymmetric bispinor:
\begin{equation}
 \epsilon_{\alpha\beta} \xi^\beta = - \xi_\alpha, \qquad
 \epsilon^{\alpha\beta} \xi_\beta = \xi^\alpha,
\end{equation} 
similarly for dotted indices. Hence, symmetry over a set on indices
implies tracelessness. This feature greatly simplifies the work with
traceless mixed symmetry tensors and spin-tensors.
The mixed symmetry tensor $ \Phi^{\mu(k),\nu(l)}$ which corresponds to
the two-row Young tableaux $Y(k,l)$ \cite{BB06} is described by a pair
of multispinors $\Phi^{\alpha(k+l)\dot\alpha(k-l)}$,
$\Phi^{\alpha(k-l)\dot\alpha(k+l)}$ in multispinor formalism; if the
tensor $\Phi^{\mu(k),\nu(l)}$ is real then:
\begin{equation}
\left(\Phi^{\alpha(k+l)\dot\alpha(k-l)}\right)^\dagger =
\Phi^{\alpha(k-l)\dot\alpha(k+l)}.
\end{equation} 
Similarly, the mixed symmetry spin-tensor $ \Psi^{\mu(k),\nu(l)}$
which corresponds to the Young tableaux $Y(k+\iz,l+\iz)$ is described
by a pair of multispinors $\Psi^{\alpha(k+l+1)\dot\alpha(k-l)}$,
$\Psi^{\alpha(k-l+1)\dot\alpha(k+l)}$; if the spin-tensor 
$\Psi^{\mu(k),\nu(l)}$ is Majorana one then
\begin{equation}
\left(\Psi^{\alpha(k+l+1)\dot\alpha(k-l)}\right)^\dagger =
\Psi^{\alpha(k-l)\dot\alpha(k+l+1)}.
\end{equation} 

In the frame-like formalism, two bases, namely the world one and the
local one are used. We denote the local basis vectors as
$e^{\alpha\dot\alpha}$; the world indices are omitted, and all the
fields are assumed differential forms with respect to them. Similarly
all the products are exterior with respect to the world indices. In
the paper, we use basis forms, i.e. antisymmetrized products of basis
vectors $e^{\alpha\dot\alpha}$. The forms are 2-form
$E^{\alpha(2)}+h.c.$, 3-form $E^{\alpha\dot\alpha}$ and 4-form $E$.
The transformation law of the forms under the hermitian conjugation
is:
\begin{equation}
(e^{\alpha\dot\alpha})^\dagger=e^{\alpha\dot\alpha}
\qquad
(E^{\alpha(2)})^\dagger=E^{\dot\alpha(2)}
\qquad
(E^{\alpha\dot\alpha})^\dagger=-E^{\alpha\dot\alpha}
\qquad
(E)^\dagger=-E
\end{equation}

\section{Equations on the parameters of superblock}

Here we provide a complete set of equations which follows from the
requirement that unfolded equations be invariant under the
supertransformations. For the supertransformations of the bosonic
sector of gauge invariant zero-forms we obtain:
\begin{eqnarray}
\label{superblock_eqs1}
\frac{\delta_{k,m}^{0+}\tilde{\beta}^{i,-}_{k+\iz,m-\iz}
+\beta^{i,+}_{k,m}\delta_{k+\iz(1+i),m-\iz(1-i)}^{0-}
-\lambda\delta_{k,m}^{i0}}{k-m}
&=&
\beta^{i,-}_{k,m}\delta_{k-\iz(1-i),m+\iz(1+i)}^{0+}
-\delta_{k,m}^{0+}\tilde{\beta}^{i,-}_{k+\iz,m-\iz}
\nonumber \\&=&
\delta_{k,m}^{0-}\tilde{\beta}^{i,+}_{k-\iz,m+\iz}-\beta^{i,+}_{k,m}\delta_{k+(1+i)\iz,m-\iz(1-i)}^{0-}
,
\nonumber \\
\frac{\delta_{k,m}^{+0}\tilde{\beta}^{-,i}_{k+\iz,m+\iz}
+\beta^{+,i}_{k,m}\delta_{k+\iz(1+i),m+\iz(1-i)}^{-0}
-\lambda\delta_{k,m}^{0i}}{k+m}
&=&
\beta^{-,i}_{k,m}\delta_{k-(1-i)\iz,m-(1+i)\iz}^{+0}-\delta_{k,m}^{+0}\tilde{\beta}^{-,i}_{k+\iz,m+\iz}
\nonumber \\&=&
\delta_{k,m}^{-0}\tilde{\beta}^{+i}_{k-\iz,m-\iz}-\beta^{+i}_{k,m}\delta_{k+\iz(1+i),m+(1+i)\iz}^{-0},
\nonumber \\
\beta^{ij}_{k,m}\delta_{k+\iz(i+j),m+\iz(i-j)}^{i0}&=&\delta_{k,m}^{i0}\tilde{\beta}^{ij}_{k+\iz i,m+\iz i},
\nonumber \\
\beta^{ij}_{k,m}\delta_{k+\iz(i+j),m+\iz(i-j)}^{0j}&=&\delta_{k,m}^{0j}\tilde{\beta}^{ij}_{k+\iz j,m-\iz j}
\end{eqnarray}
and similar conditions with inverted tildes (i.e. tilde is added above
the coefficients which do not possess one and removed from those which
have one) for the fermionic sector with half-integer $k,m$. Here $i,j$
are numbers $\pm 1$; when written as upper indices of the
coefficients, they stand for $+$ and $-$ respectively.

Similarly, for the gauge sector supertransformation parameters we get:
\begin{eqnarray}
\label{superblock_eqs2}
\frac{\gamma_{k,m}^{0+}\tilde{\alpha}^{i,-}_{k+\iz,m-\iz}
+\alpha^{i,+}_{k,m}\gamma_{k+\iz(1+i),m-\iz(1-i)}^{0-}
-\lambda\gamma_{k,m}^{i0}}{k-m}
&=&
\alpha^{i,-}_{k,m}\gamma_{k-\iz(1-i),m+\iz(1+i)}^{0+}
-\gamma_{k,m}^{0+}\tilde{\alpha}^{i,-}_{k+\iz,m-\iz}
\nonumber \\&=&
\gamma_{k,m}^{0-}\tilde{\alpha}^{i,+}_{k-\iz,m+\iz}-\alpha^{i,+}_{k,m}\gamma_{k+(1+i)\iz,m-\iz(1-i)}^{0-}
,
\nonumber \\
\frac{\gamma_{k,m}^{+0}\tilde{\alpha}^{-,i}_{k+\iz,m+\iz}
+\alpha^{+,i}_{k,m}\gamma_{k+\iz(1+i),m+\iz(1-i)}^{-0}
-\lambda\gamma_{k,m}^{0i}}{k+m}
&=&
\alpha^{-,i}_{k,m}\gamma_{k-(1-i)\iz,m-(1+i)\iz}^{+0}-\gamma_{k,m}^{+0}\tilde{\alpha}^{-,i}_{k+\iz,m+\iz}
\nonumber \\&=&
\gamma_{k,m}^{-0}\tilde{\alpha}^{+i}_{k-\iz,m-\iz}-\alpha^{+i}_{k,m}\gamma_{k+\iz(1+i),m+(1-i)\iz}^{-0},
\nonumber \\
\alpha^{ij}_{k,m}\gamma_{k+\iz(i+j),m+\iz(i-j)}^{i0}&=&\gamma_{k,m}^{i0}\tilde{\alpha}^{ij}_{k+\iz i,m+\iz i},
\nonumber \\
\alpha^{ij}_{k,m}\gamma_{k+\iz(i+j),m+\iz(i-j)}^{0j}&=&\gamma_{k,m}^{0j}\tilde{\alpha}^{ij}_{k+\iz j,m-\iz j}
\end{eqnarray}
The relations for $\tilde{\gamma}^{ij}_{k,m}$ are obtained by
inverting tildes.

\end{document}